# Review of nuclear magnetic resonance studies on iron-based superconductors*

Ma Long(马 龙) and Yu Wei-Qiang(于伟强)†

*Department of Physics, Renmin University of China, Beijing* 100872, *China*

The newly discovered iron-based superconductors have triggered renewed enormous research interest in the condensed matter physics community. Nuclear magnetic resonance (NMR) is a low-energy local probe for studying strongly correlated electrons, and particularly important for high-$T_C$ superconductors. In this paper, we review NMR studies on the structural transition, antiferromagnetic order, spin fluctuations, and superconducting properties of several iron-based high-$T_C$ superconductors, including $LaFeAsO_{1-x}F_x$, $LaFeAsO_{1-x}$, $BaFe_2As_2$, $Ba_{1-x}K_xFe_2As_2$, $Ca_{0.23}Na_{0.67}Fe_2As_2$, $BaFe_2(As_{1-x}P_x)_2$, $Ba(Fe_{1-x}Ru_x)_2As_2$, $Ba(Fe_{1-x}Co_x)_2As_2$, $Li_{1+x}FeAs$, $LiFe_{1-x}Co_xAs$, NaFeAs, $NaFe_{1-x}Co_xAs$, $K_yFe_{2-x}Se_2$, and $(Tl,Rb)_yFe_{2-x}Se_2$.

**Keywords:** iron-based superconductors, nuclear magnetic resonance (NMR), spin fluctuation, superconductivity

**PACS:** 74.70.Xa, 74.25.nj, 76.60.–k

## 1. Introduction

The discovery of high-$T_C$ superconductivity (SC) in iron pnictide materials[1–4] stimulates large research enthusiasm on unconventional superconductors. Up to now, several families of iron-based superconductors sharing a similar phase diagram have been found.[1,5–7] The parent compounds are metallic antiferromagnets. Via carrier doping or under pressure, the antiferromagnetism (AFM) was gradually suppressed and SC was developed. With time going by, it has been found that the correlation effect among the lattice structure, magnetism, spin fluctuations, and superconductivity are important for understanding the properties and the paring mechanism of the superconductivity.

Nuclear magnetic resonance (NMR) detects the on-site hyperfine field originating from the electronic bath. Therefore, it is very sensitive to static electronic spin susceptibility, magnetic order, and low-energy magnetic excitations. In this paper, we review NMR studies on several iron-based superconductors. We first introduce principles of NMR in Section 2. In Section 3, we focus on NMR studies of the paring symmetry of several compounds, including $LaFeAsO_{1-x}F_x$, $LaFeAsO_{0.7}$, $KFe_2As_2$, $BaFe_2(As_{0.67}P_{0.33})_2$, and $K_yFe_{2-x}Se_2$. In Section 4, we review the NMR study of the structural transition in $BaFe_2As_2$ and NaFeAs. In Section 5, we concentrate on the NMR studies on the long-range ordered AFM (LROAFM) in $BaFe_2As_2$ and NaFeAs. The competition between SC and LROAFM as well as their coexistence in several underdoped samples, including $BaFe_2(As_{0.75}P_{0.25})_2$, $Ba(Fe_{0.77}Ru_{0.23})_2As_2$, and $Ba_{0.77}K_{0.23}Fe_2As_2$, are reviewed in Section 6. In Section 7, we review the studies of the spin fluctuations in the normal states of $LaFeAsO_{1-x}F_x$, $Ba(Fe_{1-x}Co_x)_2As_2$, $BaFe_2(As_{1-x}P_x)_2$, $Ba_{1-x}K_xFe_2As_2$, $Ca_{0.33}Na_{0.67}Fe_2As_2$, $Li_{1+x}FeAs$, $LiFe_{1-x}Co_xAs$, $NaFe_{1-x}Co_xAs$, $K_yFe_{2-x}Se_2$, $(Tl,Rb)_yFe_{2-x}Se_2$, and $BaFe_2As_2$, where two types of spin fluctuations, one from Fermi surface nesting and one from local spins, are suggested.

## 2. Principles of NMR[8–10]

### 2.1. NMR Hamiltonians

The nuclear spin system in condensed matter can be considered to exist in the thermal bath of the electron system via hyperfine coupling. The typical energy scale of nuclear spins is about three orders smaller than that of electron system. As a result, nuclear magnetic resonance is very suitable for the study of static magnetism and low-energy spin fluctuations in condensed matter. The interactions between electrons and nuclei can be described with the Hamiltonian, given by

$$H = H_{\text{Zeeman}} + H_{\text{e-n}} + H_{\text{eQ}} + H_{\text{dip}}. \quad (1)$$

The first term on the right-hand side describes the Zeeman splitting between the degenerate nuclear energy levels when an external field $\boldsymbol{H}_{\text{ext}}$ is applied. The detailed Hamiltonian form can be written as

$$H_{\text{Zeeman}} = \gamma\hbar \boldsymbol{H}_{\text{ext}} \cdot \boldsymbol{I}, \quad (2)$$

*Project supported by the National Natural Science Foundation of China (Grant Nos. 11074304 and 11222433) and the National Basic Research Program of China (Grant Nos. 2010CB923004 and 2011CBA00112).

†Corresponding author. E-mail: wqyu_phy@ruc.edu.cn

where $\gamma$ and $\boldsymbol{I}$ are the nuclear gyromagnetic ratio and spin, respectively. In the common NMR experiments, this Zeeman splitting energy is the largest among the four terms for the nuclear system, and the other three terms of the Hamiltonian can be treated as perturbations.

The second term on the right-hand side of Eq. (1) is given by

$$H_{\mathrm{e-n}} = \gamma\hbar\boldsymbol{I}\cdot\sum_i(2\mu_{\mathrm{B}})\left(\frac{\boldsymbol{l}_i}{r_i^3}-\frac{\boldsymbol{S}_i}{r_i^3}+3\frac{\boldsymbol{r}_i(\boldsymbol{S}_i\cdot\boldsymbol{r}_i)}{r_i^5}+\frac{8}{3}\pi\boldsymbol{S}_i\delta(\boldsymbol{r}_i)\right). \quad (3)$$

The term $H_{\mathrm{e-n}}$ is the detailed form for the hyperfine interaction between the nuclear system and the electric bath. The first term in Eq. (3) represents the interactions between the nuclei and orbital motion of electrons, which contributes to a chemical shift in the spectral frequency. The second and third term describe the dipolar interactions between nuclei and electrons. The last term describes onsite Fermi contact interaction due to the overlapped wave functions of nuclei and electrons. The hyperfine interaction is the main channel for the spin-lattice relaxation process on the time scale $T_1$, which gives the information about the low-energy excitations in the electron system, which is important for the study of condensed matter.

The third term on the right-hand side of Eq. (1) is written as

$$H_{\mathrm{eQ}} = \frac{eQV_{zz}}{4I(2I-1)}[(3{I_z}^2-I^2)+\eta({I_x}^2-{I_y}^2)]. \quad (4)$$

The term $H_{\mathrm{eQ}}$ is the nuclear quadrupole interactions with the crystalline electronic field gradient (EFG) as shown in Eq. (3), where $Q$ is the nuclear quadrupole moment, $V_{\alpha\beta}$ is the component of the EFG tensor, and $\eta$ is the asymmetry parameter defined as $\eta = \mid V_{xx}-V_{yy}\mid/V_{zz}$. The quadrupolar interaction is non-zero only for the nuclei with $I>1/2$, which gives the important information about the lattice structure reflected by EFG. In fact, this interaction also breaks the degeneracy of the nuclear energy levels, and the nuclear quadrupole resonance (NQR) technique can be performed under zero field.

The last term on the right-hand side of Eq. (1) is written as

$$H_{\mathrm{dip}} = \frac{1}{2}\sum_{j,k=1}^{N}\left(\frac{\boldsymbol{\mu}_j\cdot\boldsymbol{\mu}_k}{r_{jk}^3}-\frac{3(\boldsymbol{\mu}_j\cdot\mathbf{r}_{jk})(\boldsymbol{\mu}_k\cdot\mathbf{r}_{jk})}{r_{jk}^5}\right). \quad (5)$$

This term is the dipolar interaction between each nuclei with moments $\boldsymbol{\mu}$ and a distance of $\boldsymbol{r}_{jk}$, which describes the energy transfer within the nuclear system. Usually, the dipolar interaction contributes to the nuclear diffusing mechanism with the typical time scale $T_2$ known as the spin–spin relaxation time and a static broadening of the spectra for type-II superconductors. The spin–spin relaxation rate $1/T_2$ gives the detailed information about the vortex dynamics.

### 2.2. NMR spectra

The NMR spectrum is mainly determined by the first two parts of the Hamiltonian given by Eq. (1) with an effective magnetic field composed of the applied and internal hyperfine field. Here we treat nuclei with $I=3/2$ under crystalline EFG without in-plane asymmetry. Under a high magnetic field limit, the nuclear quadrupole interactions with crystalline EFG shown by Eq. (4) can be treated with the perturbation theory. Various energy levels are written as

$$E_m = E_m^{(0)}+E_m^{(1)}+E_m^{(2)}, \quad (6)$$

where the energy level without perturbation $E_m^{(0)}$, the first-order and second-order perturbations to the Zeeman levels $E_m^{(1)}$ and $E_m^{(2)}$ are given by

$$E_m^{(0)} = -\gamma\hbar H_{\mathrm{eff}}m = -h\nu_{\mathrm{L}}m, \quad (7)$$

$$E_m^{(1)} = \frac{1}{4}h\nu_Q(3\mu^2-1)(m^2-\frac{1}{3}a), \quad (8)$$

$$E_m^{(2)} = -h\left(\frac{\nu_Q^2}{12\nu_{\mathrm{L}}}\right)m\left[\frac{3}{2}\mu^2(1-\mu^2)(8m^2-4a+1)+\frac{3}{8}(1-\mu^2)^2(-2m^2+2a-1)\right], \quad (9)$$

where $a=I(I+1)$, $\nu_{\mathrm{Q}}=3e^2qQ/[2hI(2I-1)]$, $\mu=\cos\theta$, and $\nu_{\mathrm{L}}=\gamma H/2\pi$.

Based on these calculations, $^{75}$As resonance peaks in iron pnictides are shown schematically in Fig. 1. With the nuclear quadrupole interaction with crystalline EFG present in the iron pnictide, the first-order correction results in $^{75}$As spectra with a single central line ($\nu_0$) and two satellites ($\nu_1$ and $\nu_2$). The angular dependences of the satellite frequencies are given by $\nu_1=\nu_{\mathrm{L}}+\nu_{-1/2}^{(1)}$ and $\nu_2=\nu_{\mathrm{L}}+\nu_{3/2}^{(1)}$, where $\nu_{\mathrm{L}}$, $\nu_{-1/2}^{(1)}$, and $\nu_{3/2}^{(1)}$ are given by

$$\nu_{\mathrm{L}} = \frac{E_{(m-1)}^{(0)}-E_m^{(0)}}{h} = \gamma H_{\mathrm{eff}}, \quad (10)$$

$$\nu_{-1/2}^{(1)} = \frac{E_{-3/2}^{(1)}-E_{-1/2}^{(1)}}{h} = \nu_{\mathrm{Q}}\frac{3\mu^2-1}{2}, \quad (11)$$

$$\nu_{3/2}^{(1)} = \frac{E_{1/2}^{(1)}-E_{3/2}^{(1)}}{h} = -\nu_{\mathrm{Q}}\frac{3\mu^2-1}{2}. \quad (12)$$

The second-order correction causes a spectral shift of the central line, $\nu_3=\nu_{\mathrm{L}}+\nu_{1/2}^{(2)}$, whose angular dependence is given by

$$\nu_{1/2}^{(2)} = \frac{E_{-1/2}^{(1)}-E_{1/2}^{(1)}}{h} = \frac{-3\nu_{\mathrm{Q}}^2}{16\nu_{\mathrm{L}}}(1-\mu^2)(9\mu^2-1). \quad (13)$$

For the nuclei with $I$=3/2, the second-order correction will not shift the satellites accidentally.

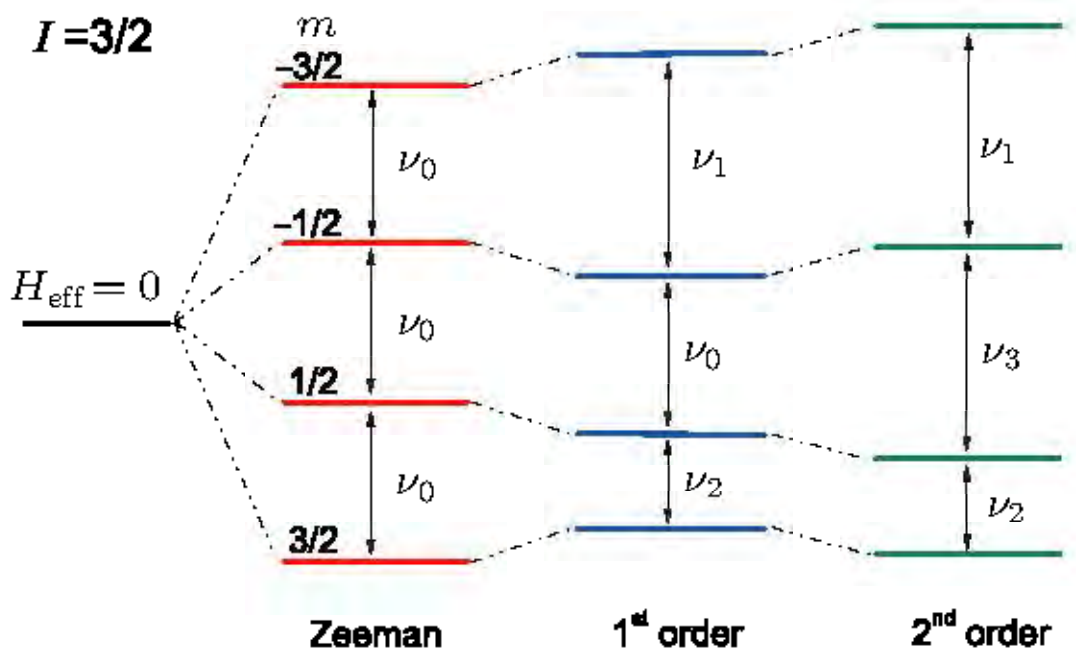

**Fig. 1.** Schematic diagram of the nuclear energy levels under the magnetic field and crystalline EFG.

### 2.3. NMR Knight shift and the spin-lattice relaxation rate

In a simple metal, the dominant interaction between nuclear spin and quasiparticle spin is the onsite Fermi contact hyperfine interaction, namely,

$$H_{\rm hf}=A\boldsymbol{I}\cdot\boldsymbol{S}=AI_zS_z+\frac{1}{2}A(I_+S_-+I_-S_+). \tag{14}$$

The first diagonal term contributes to a spectral shift from the resonance frequency of isolated nuclear spins, known as Knight shift in the form of

$$K=\frac{\nu-\nu_{\rm L}}{\nu_{\rm L}}, \tag{15}$$

where $\nu$ and $\nu_{\rm L}$ denote the observed resonance frequency and Lamor frequency of isolated nuclear spins under the magnetic field, respectively. The spin part of the Knight shift follows $K_{\rm S}=A\langle S_z\rangle/(\gamma_{\rm n}\hbar H_{\rm ext})=A\chi_{\rm S}/(\gamma_{\rm n}\hbar\mu_{\rm B}N_{\rm A})\equiv A_{\rm hf}\chi_{\rm S}/(\mu_{\rm B}N_{\rm A})$, which is proportional to the time average of quasiparticle spin polarization $\langle S_z\rangle$ due to the applied magnetic field. In addition, $\chi_{\rm S}$, $\mu_{\rm B}$, and $N_{\rm A}$, respectively, denote the quasiparicle spin susceptibility, Bohr magneton, and Avogadro's number. As NMR is a local probe, the Knight shift measures the intrinsic spin susceptibility of the electronic system.

In a simple metal, this susceptibility is related to the density of state near Fermi surface $N(E_{\rm F})$, and $\chi_{\rm S}=\mu_{\rm B}^2N(E_{\rm F})N_{\rm A}$. As a result, Knight shift in a simple metal can be expressed as

$$K_{\rm S}(T)=A_{\rm hf}\mu_{\rm B}N(E_{\rm F}). \tag{16}$$

The second off-diagonal term in Eq. (14) describes the spin transfer between the nucleus and quasiparticle system, which gives detailed information about spin dynamics of electronic system. As the gap between nuclear Zeeman levels is about three orders of magnitude smaller than the energy scale of quasiparticle excitations in condensed matter, the spin-lattice relaxation rate, $1/T_1$, is written as

$$\frac{1}{T_1}=\gamma_{\rm n}^2A_{\rm hf}^2\int_0^\infty\langle S_+(t)S_-(0)\rangle\,{\rm e}^{{\rm i}\omega_{\rm L}t}\,{\rm d}t, \tag{17}$$

based on the perturbation theory. It measures a summarization of low-energy elementary excitations at different $q$ modes. However, these low-energy excitations are of great importance for determining the property of condensed matter. By employing the fluctuation–dissipation theorem, equation (17) can be generally expressed as a summarization of the dynamical susceptibility $\chi(q,\omega_{\rm L})$,

$$\frac{1}{T_1}=\frac{\gamma_{\rm n}^2k_{\rm B}T}{\mu_{\rm B}^2}\sum_q A(q)^2\frac{\chi''(q,\omega_{\rm L})}{\omega_{\rm L}}, \tag{18}$$

which is very useful for the analysis of the magnetic correlations in the electronic system.

In a simple metal without electronic correlations, electron–hole pair excitation is the main contribution to the spin-lattice relaxation rate. Thus, we have

$$\frac{1}{T_1}=\frac{\gamma_{\rm n}^2A_{\rm hf}^2}{2}\int_0^\infty N(E_{\rm i})N(E_{\rm f})f(E_{\rm i})(1-f(E_{\rm f}))\,{\rm d}E_{\rm i}, \tag{19}$$

where $E_{\rm i}$ and $E_{\rm f}$ are energies of the initial and final state, respectively. As the energy gap between $E_{\rm i}$ and $E_{\rm f}$ is the Zeeman splitting energy, which is very small compared with the energy scale of electronic system, $N(E_{\rm i})$ and $N(E_{\rm f})$ can be written as density of states near Fermi surface, $N(E_{\rm F})$. As a result,

$$\frac{1}{T_1(T)}=\pi\hbar\gamma_{\rm n}^2A_{\rm hf}^2N(E_{\rm F})^2k_{\rm B}T. \tag{20}$$

As the temperature-independent Knight shift $K$ is proportional to $N(E_{\rm F})$, we can essentially obtain a constant $T_1TK^2$ for each nucleus without influence of the quasiparticle excitation, which is known as the Korringa relation given by

$$T_1TK^2=\frac{\mu_{\rm B}^2}{\pi k_{\rm B}\hbar\gamma_{\rm n}^2}\equiv S_0. \tag{21}$$

From Eq. (20), the Korringa constant $S_0$ is defined, which is a measure for the electronic correlations in condensed matter physics. However, for a system with electronic correlation, the Korringa relation is not usually followed, as will be shown later.

## 3. NMR study on the paring symmetry of iron-based superconductors

The Knight shift and spin-lattice relaxation rate give important information about the Cooper-pair spin state and superconducting gap symmetry. For the spin part, the Knight shift determines singlet ($S=0$) or triplet ($S=1$) paring, depending on whether $K_{\rm S}(T)$ drops below $T_{\rm C}$ with the field applied in all directions.

In this section, we give a review on the NMR study on the paring symmetries of iron-based superconductors. The singlet paring is always indicated by the sharp drop of Knight shift below $T_{\rm C}$ in all directions. However, unconventional paring symmetry for the orbital part is also suggested in iron-based superconductors, owing to the absence of the coherence peak and the low power law of the spin-lattice relaxation rate as shown below.

The temperature dependences of the Knight shift in $LaFeAsO_{0.89}F_{0.11}$ and $LaFeAsO_{0.7}$ samples reported by Kawabata *et al.*[11] and Terasaki*et al.*[12] are shown in Fig. 2. In Fig. 2, the temperature-independent orbital contribution has been eliminated. Above $T_C$, the Knight shift shows weak temperature dependence. With temperature decreasing below $T_C$, the sharp drop of the Knight shift to zero is observed, indicating the spin susceptibility decreasing in the superconducting state. This drop provides strong evidence for the spin singlet paring state.

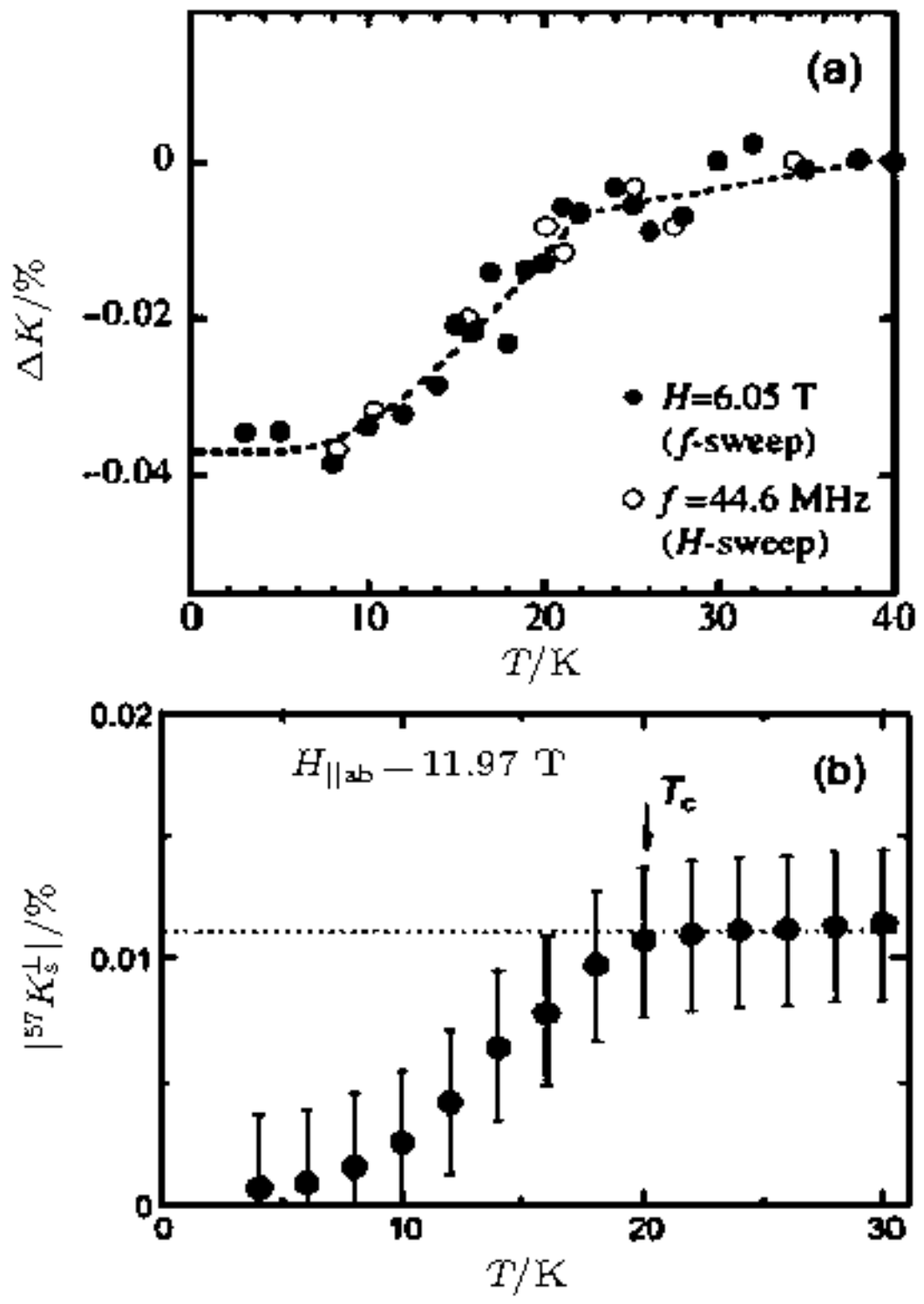

**Fig. 2.** (a) $^{75}$As Knight shift in $LaFeAsO_{0.89}F_{0.11}$ as a function of temperature, normalized by its value at $T = 40$ K.[11] (b) The temperature dependence of the spin part of $^{57}$Fe Knight shift $|^{57}K_s^{\perp}|$ in $LaFeAsO_{0.7}$.[12]

For fully gapped SC, $1/(T_1T)$ usually has a coherence peak. Particularly for s-wave paring, $1/(T_1T)$ shows a thermal activation behavior $(1/(T_1T) \sim \exp(-\frac{\Delta}{k_B T_C}))$ at very low temperatures. For superconductors with line node, $1/(T_1T)$ follows a power-law behavior, $1/(T_1T) \sim T^3$, far below $T_C$. The spin-lattice relaxation rate measures the effective low-energy elementary excitations, and thus is a good probe for the paring symmetry of the orbital part.

Oka *et al.*[13] reported on their NQR results of the superconducting $LaFeAsO_{1-x}F_x$ materials. With the NQR experiment, they eliminated the vortex core contributions to the spin-lattice relaxation rates. Figure 3 shows doping evolutions of the temperature dependence of $1/T_1$. Hamp structures in $1/T_1$ are observed for $x = 0.06$–$0.1$ samples with $T \sim 0.4T_C$. At lower temperatures, $1/T_1$ decreases sharper. However, $1/T_1$ in the low temperature region decreases less steeply with the increase in doping level. In the $x = 0.15$ sample, the temperature dependence of $1/T_1$ is close to $T^3$, instead of the emergence of the hump structure. Oka *et al.*[13] proposed that the marked hump structure is consistent with the double superconducting gap opened on the different Fermi surface due to the sign reversing s-wave ($s^{\pm}$) gap symmetry from interband transitions of electron and hole pocket. The $1/T_1$ data can be reproduced with $s^{\pm}$ gap symmetry with impurity scattering, namely,

$$\frac{T_1(T_C)}{T_1(T)}\frac{T_C}{T} = \frac{1}{4T}\int_{-\infty}^{+\infty}\frac{d\omega}{\cosh^2(\omega/2T)}(W_{GG}+W_{FF}),$$

where

$$W_{GG} = [\langle \mathrm{Re}\{(\omega+i\eta)/\sqrt{(\omega+i\eta)^2+|\Delta(k_F)|^2}\}\rangle_{k_F}]^2,$$
$$W_{FF} = [\langle \mathrm{Re}\{1/\sqrt{(\omega+i\eta)^2+|\Delta(k_F)|^2}\}\Delta(k_F)\rangle_{k_F}]^2.$$

The parameter $\eta$ measures the strength of impurity scattering. When the doping level is increased to $x = 0.15$, the temperature dependence is very close to $T^3$, which is consistent with $s^{\pm}$ gap symmetry with strong impurity scattering.

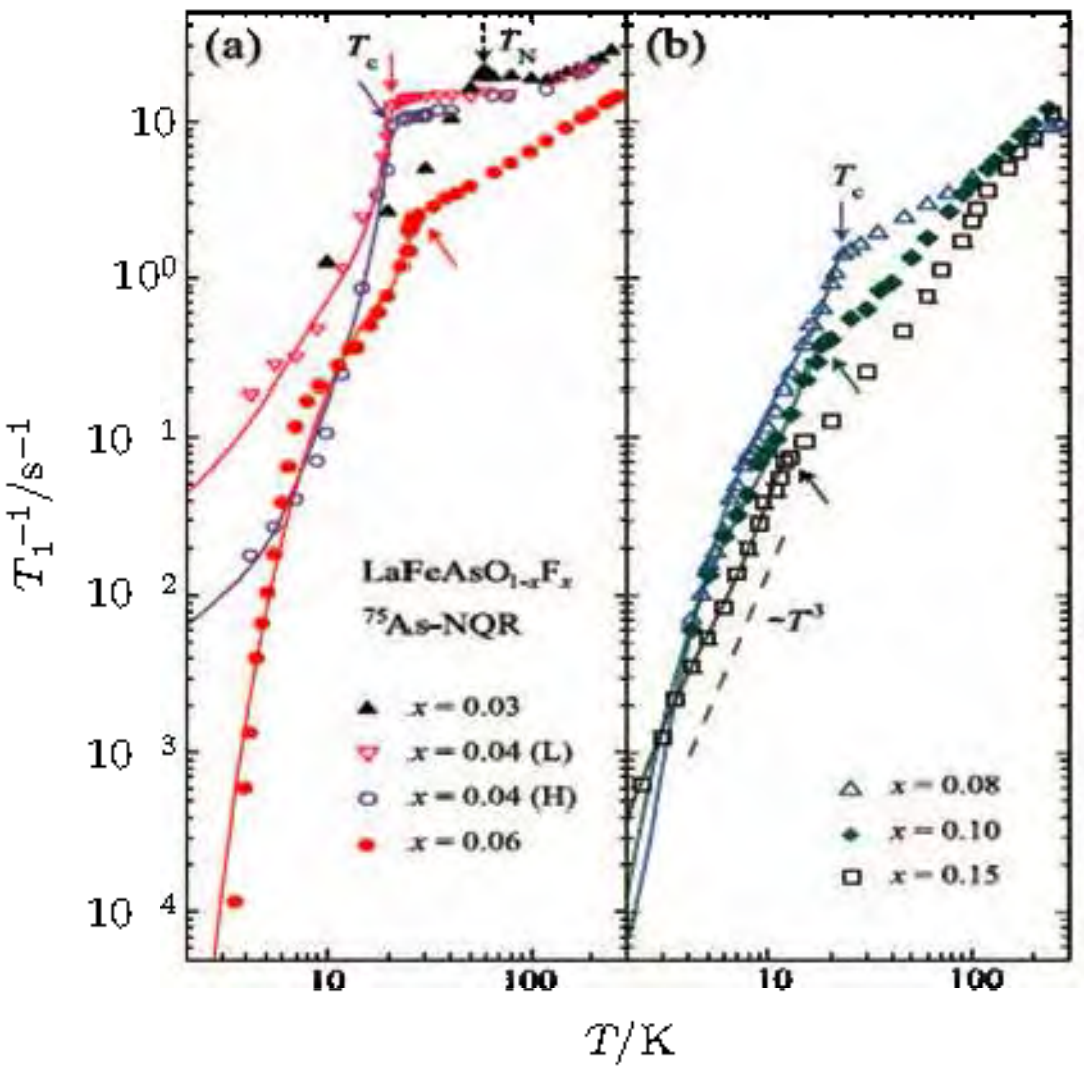

**Fig. 3.** $^{75}$As spin-lattice relaxation rate $1/T_1$ in $LaFeAsO_{1-x}F_x$ with (a) $x = 0.03$, 0.04, and 0.06, (b) $x = 0.08$, 0.10, and 0.15. The solid lines show the fittings to the $s^{\pm}$ superconducting gap model with impurity scattering. The power law behavior, $1/T_1 \propto T^3$, is shown by the dashed line.[13]

However, nodal excitations have also been proposed in heavily overdoped $KFe_2As_2$. Fukazawa *et al.*[14] reported on their measurements of $1/T_1$ in $KFe_2As_2$ sample using NQR technique as shown in Fig. 4. Below $T_C = 3.5$ K, no coherence peak is seen, suggesting the unconventional SC in $KFe_2As_2$. Between $T = 3.5$ K and 0.6 K, $1/T_1$ follows $T^{1.4}$ dependence, indicating strong low-energy quasi-particle excitations in the superconducting state. This low power-law behavior can be explained with multiple gaps with line node, which is also proved by heat capacity and heat transport measurements.[14,15]

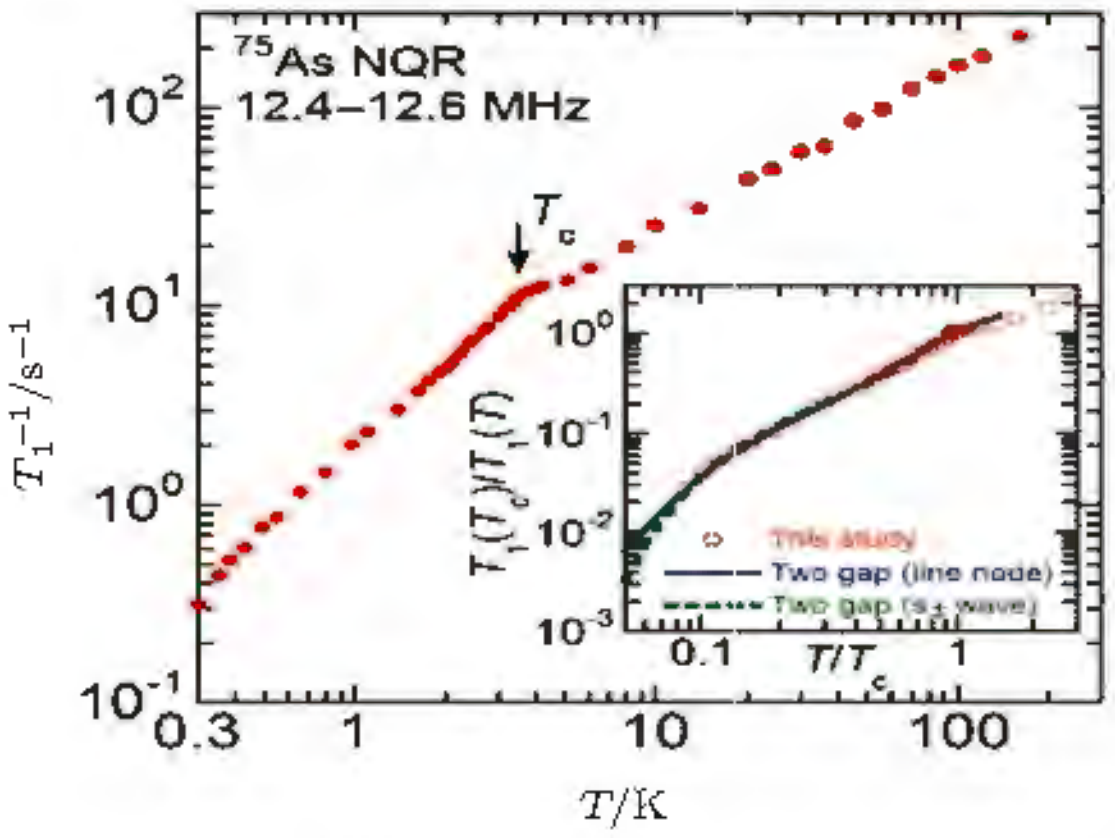

**Fig. 4.** The temperature dependence of $1/T_1$ in the heavily hole-doped $KFe_2As_2$ is shown on the logarithmic axes to compare with the power law behavior.[14]

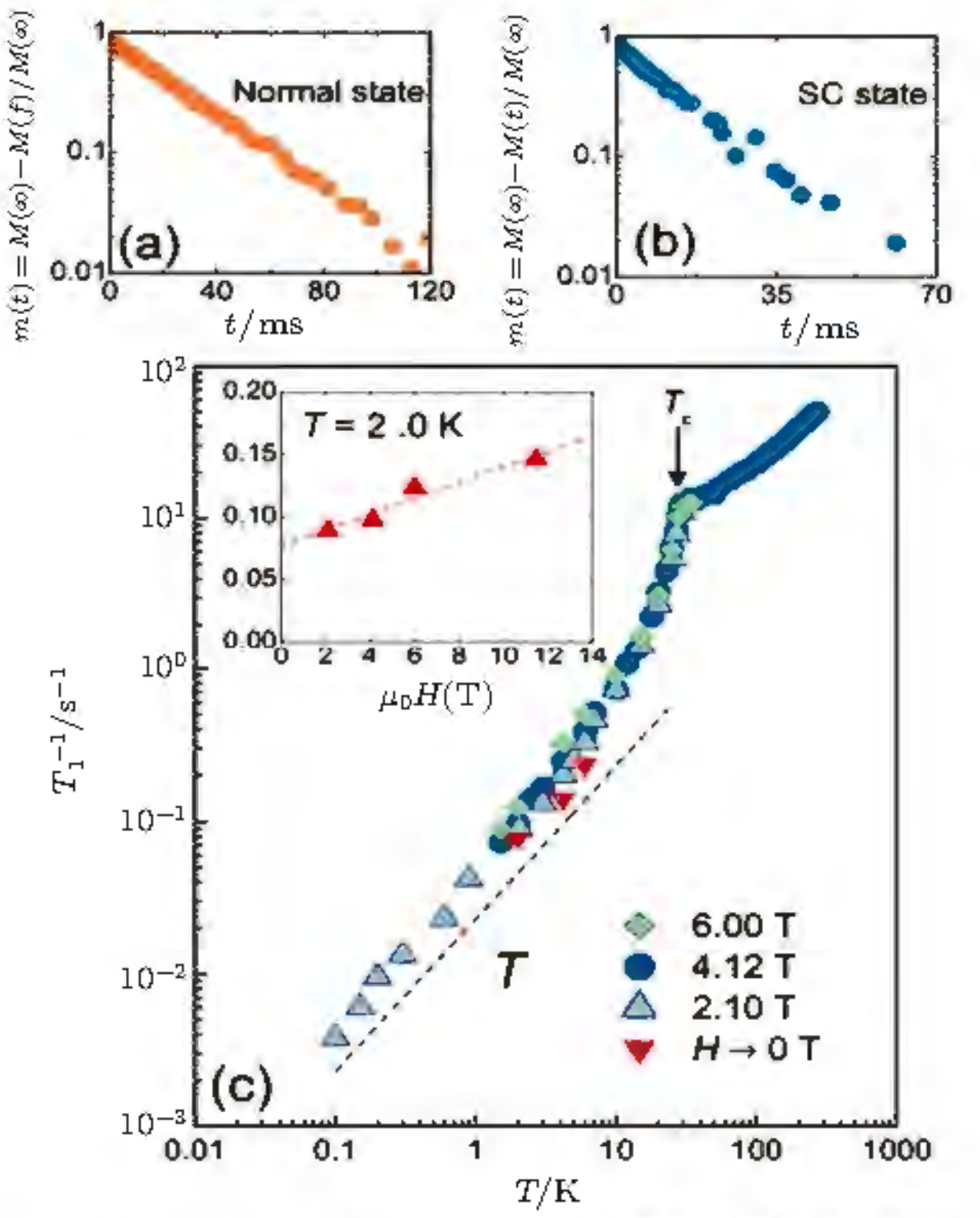

**Fig. 5.** $^{31}P$ magnetization recovery curves in $BaFe_2(As_{0.67}P_{0.33})_2$ under a field of 4.12 T at (a) $T = 175$ K and (b) 1.5 K. (c) Plots of $^{31}P$ $1/T_1$ versus temperature under different field intensities. The field dependence of $1/T_1$ at $T = 2$ K is shown in the inset of panel (c). With the dashed line, $1/T_1$ at $H = 0$ is extrapolated.[16]

Strong nodal excitations are also proposed in the phosphorus-doped $BaFe_2(As_{1-x}P_x)_2$. As shown in Fig. 5, Nakai *et al.*[16] reported on their $1/T_1$ data of $BaFe_2(As_{0.67}P_{0.33})_2$ polycrystalline samples. Below $T_C$, $1/T_1$ of $^{31}P$ drops abruptly and no evidence for coherence peak is observed. Between 4 K and 100 mK, $1/T_1$ shows linear temperature dependence, indicating strong low-energy quasi-particle excitations in the superconducting state. For $T \le 10$ K, $1/T_1$ show clear field dependence (see the inset in Fig. 5(c)). By extrapolating $1/T_1$ to zero field, the $T_1T$=constant behavior is also observed below 4 K. As a result, the existence of strong excitations in the superconducting state at $H \to 0$ indicates the existence of gap nodes of the superconducting state instead of the vortex contributions, which is consistent with the thermal-conductivity results.[17] The later ARPES study reveals a circular line node on the largest hole Fermi surface around the $Z$ point at the Brillouin zone boundary under the s$^{\pm}$ paring symmetry.[18]

The newly discovered ternary iron selenide $A_yFe_{2-x}Se_2$ ($A$ = K, Rb, Cs, ...)[20] does not have the hole pocket as reported by ARPES.[21] This raises a question of whether $K_yFe_{2-x}Se_2$ has a similar paring symmetry to iron pnictides. Figure 6 displays the Knight shift data of the newly discovered $K_yFe_{2-x}Se_2$ reported by Yu *et al.*[19] Above $T_C \sim 30$ K, the Knight shift $^{77}K$ increases with temperature in the manner of $^{77}K = a + bT^2$, which indicates strong local spin fluctuations and will be discussed later. In the case of temperatures below $T_C$, $^{77}K$ drops sharply with field along both $ab$ and $c$ directions, providing direct evidence for spin singlet SC in $K_yFe_{2-x}Se_2$. The superconducting diamagnetic contribution to $K_S$ below $T_C$ is ruled out in a later report about $(Tl,Rb)_yFe_{2-x}Se_2$.[22]

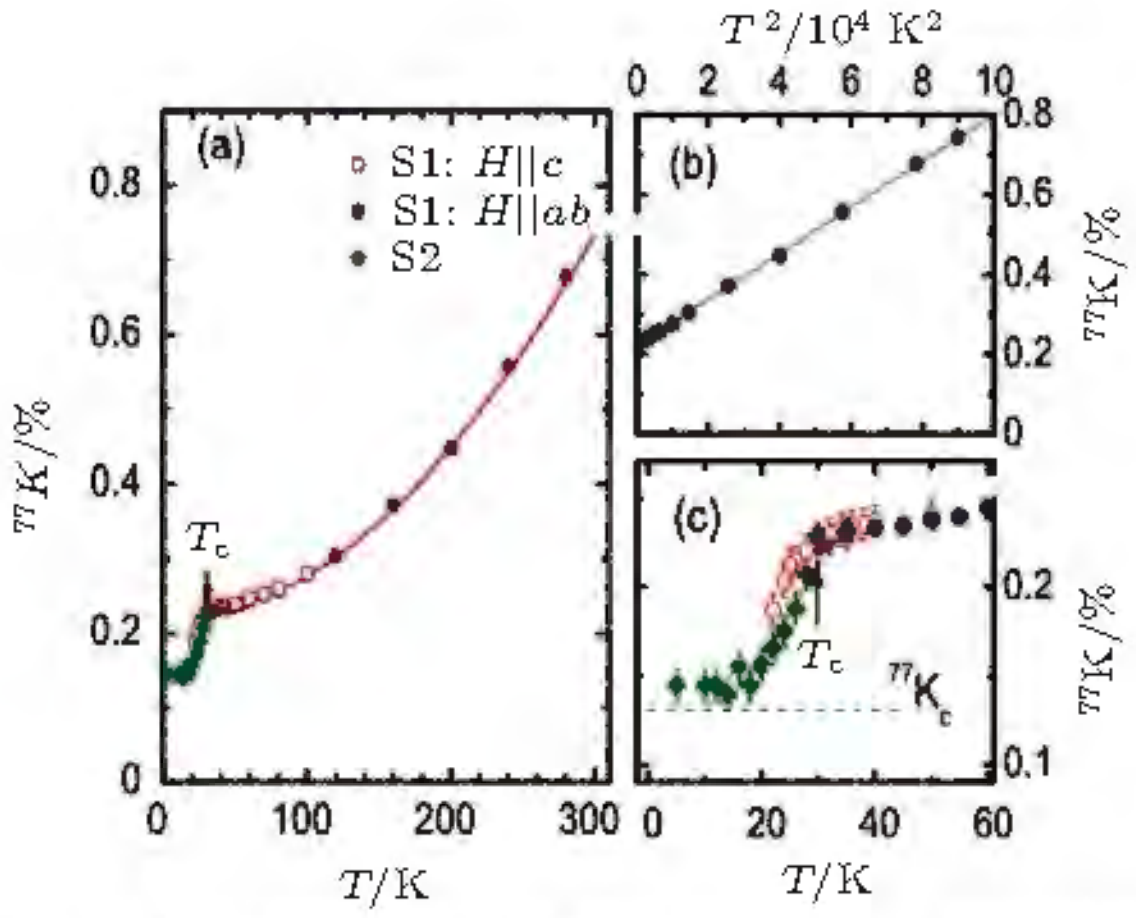

**Fig. 6.** (a) $^{77}Se$ Knight shift in $K_yFe_{2-x}Se_2$. Solid line: the quadratic temperature dependence of $^{77}K$. (b) $^{77}K$ versus $T^2$. Solid line: the perfect fitting to the function $K(T) = a + bT^2$. (c) The temperature dependence of $^{77}K$ in the low-temperature region. The orbital contribution to the Knight shift $K_c$ is denoted by the dotted horizontal line.[19]

The spin-lattice relaxation rate of $^{77}Se$ is shown in Fig. 7 to further study the orbital symmetry.[19] Above $T_C$, $1/^{77}T_1$ increases with increasing temperature due to the existence of strong local spin fluctuations. When temperature decreases below $T_C$, $1/^{77}T_1$ drops sharply and shows $T^6$ temperature dependence without any evidence for coherence peak. Below $T_C/2$, $1/^{77}T_1$ shows $T^2$ temperature dependence, which may be caused by the contribution from the localized quasi-particle excitations from vortex cores due to strong NMR field. The thermal activation function, $1/T_1 \sim \exp(-\Delta/k_BT)$, with $\Delta$ determined to be 10.3 meV from ARPES measurements,[21] works well for the $1/^{77}T_1$ data in the temperature region of $T_C > T > T_C/2$, and therefore supports a fully gapped superconductor.

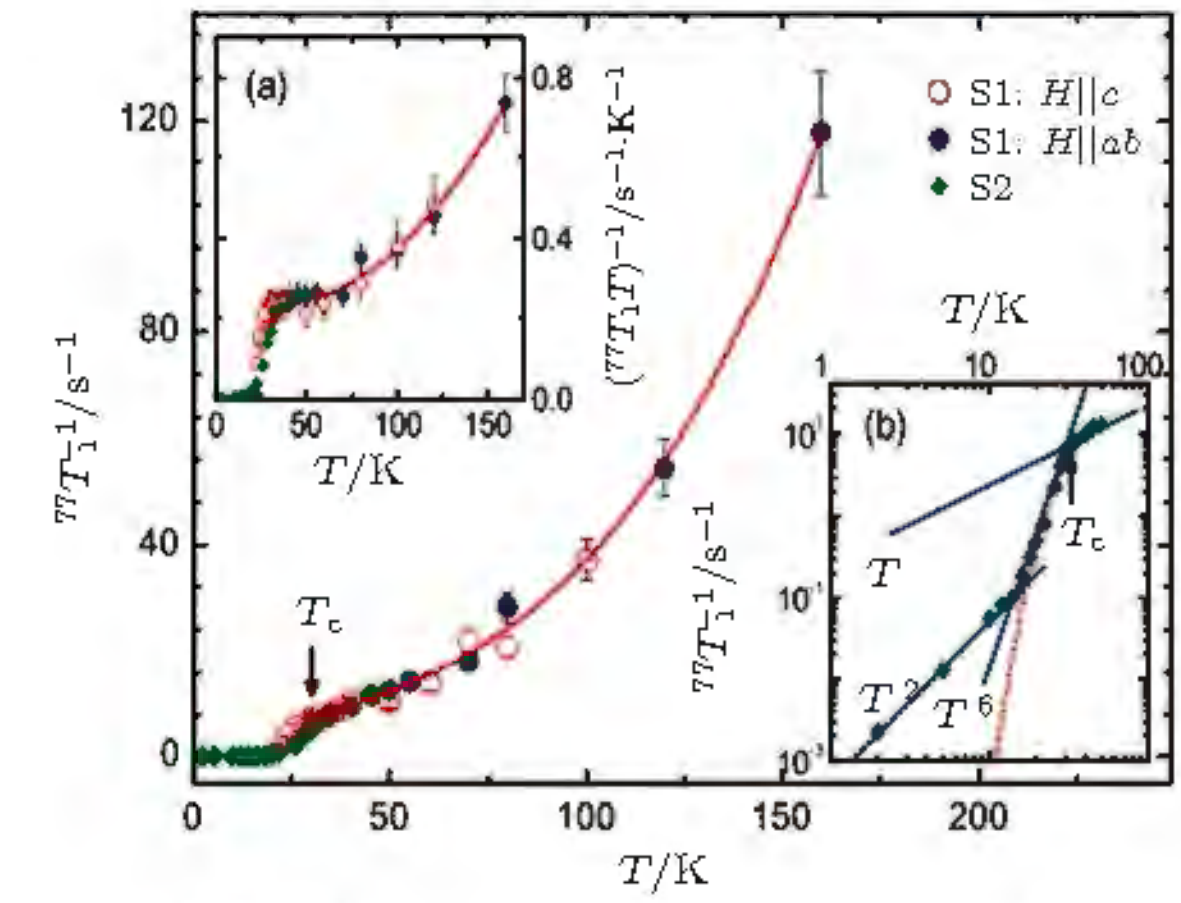

**Fig. 7.** Plots of $^{77}$Se spin-lattice relaxation rate versus temperature in $K_yFe_{2-x}Se_2$. Inset (a) shows the temperature dependence of $1/(^{77}T_1T)$ and inset (b) displays $1/^{77}T_1$ in the low-temperature region to analyze the orbital symmetry of the superconductivity. The power law ($1/T_1 \sim T^n$) and thermal activation temperature dependence ($1/^{77}T_1 = A\exp(-\Delta/k_BT)$) are shown, respectively, by the solid and dotted lines.[19]

The absence of coherence peak in iron pnictides is always attributed to the $s^{\pm}$ paring symmetry on the separate Fermi surface.[23–26] However, this interpretation is not applicable to $K_yFe_{2-y}Se_2$ system, owing to the absence of hole pocket at the center of the Brillouin zone from ARPES investigations.[21] The possibility that the coherence peak may be suppressed by the strong applied field is ruled out by the low field NMR study on the counterpart $Tl_{0.47}Rb_{0.34}Fe_{1.63}Se_2$.[22] The reason for the absence of the coherence peak still needs further studies, which may shed new light on the paring properties of electrons in iron-based superconductors.

## 4. Structural transition in iron pnictides

In the parent compounds of iron pnictides, the tetragonal to orthorhombic phase transitions are always observed prior to or during the antiferromagnetic transition. In this section, we review the NMR results on the correlations between structural transition and the LROAFM in $BaFe_2As_2$ and NaFeAs. The structural transitions in iron pnictides are strongly coupled to the magnetic correlations.

Figure 8 displays the temperature dependence of the quadrupole resonance frequency $\nu_q$ in $BaFe_2As_2$, reported by Kitagawa *et al.*[27] As mentioned in Section 2, $\nu_q$ directly measures the quadrupole interaction between the nuclear quadrupole moment and crystalline EFG, which is determined by the lattice structure. With temperature decreasing, the $\nu_q^c$ gradually decreases and shows a sudden jump at $T_S \sim 135$ K, indicating the structural transition. Correspondingly, the asymmetry parameter increases abruptly from zero in the tetragonal phase to a finite value below the transition temperature, consistent with the tetragonal-to-orthorhombic transition.

The temperature dependence of the asymmetry parameter and $\nu_q^c$ have important implications for the structural transition. The relation, $|\nu_q^a - \nu_q^b|/|\nu_q^c| > 1$, indicates that the principal axis of EFG has been rotated by 90° across the transition. Compared with the little difference between $a$ and $b$ axis, a large anisotropy of the band structure or the Fe–As bonds is suggested in the $ab$ plane.

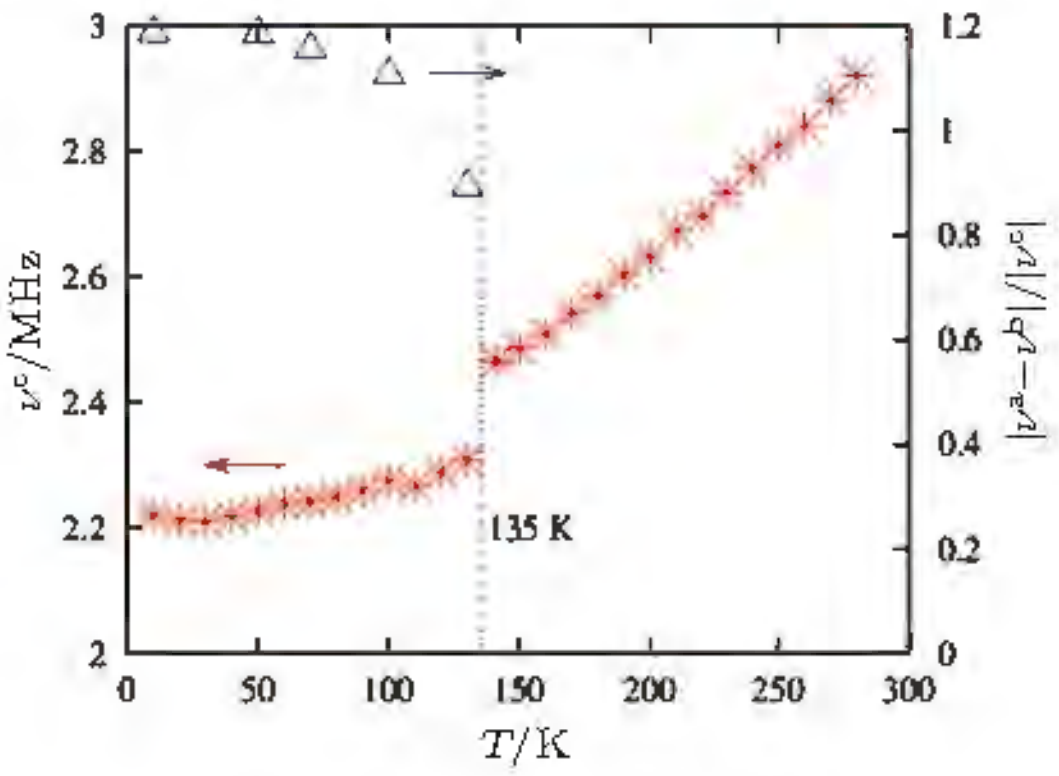

**Fig. 8.** Temperature dependence of the quadrupole splitting frequency with the field applied along the crystalline $c$ axis, and the in-plane EFG asymmetry parameter in $BaFe_2As_2$.[27]

In order to study the driving force for the structural transition, Ma *et al.*[28] reported on their NMR results of NaFeAs single crystals, where the structural transition separates from antiferromagnetic transition. The structural transition is monitored by line splitting of the $^{75}$As NMR satellite spectrum shown in Fig. 9(b). The lattice twinning is observed due to the in-plane anisotropy of $^{75}\nu_q$, with field along the $a$ or $b$ axis of the orthorhombic phase, as shown below 55 K in the inset of Fig. 9(b). The antiferromagnetic transition is also confirmed from the line splitting of $^{75}$As central peak below 40.5 K with an in-plane magnetic field, as shown in Fig. 9(a), which will be discussed later.

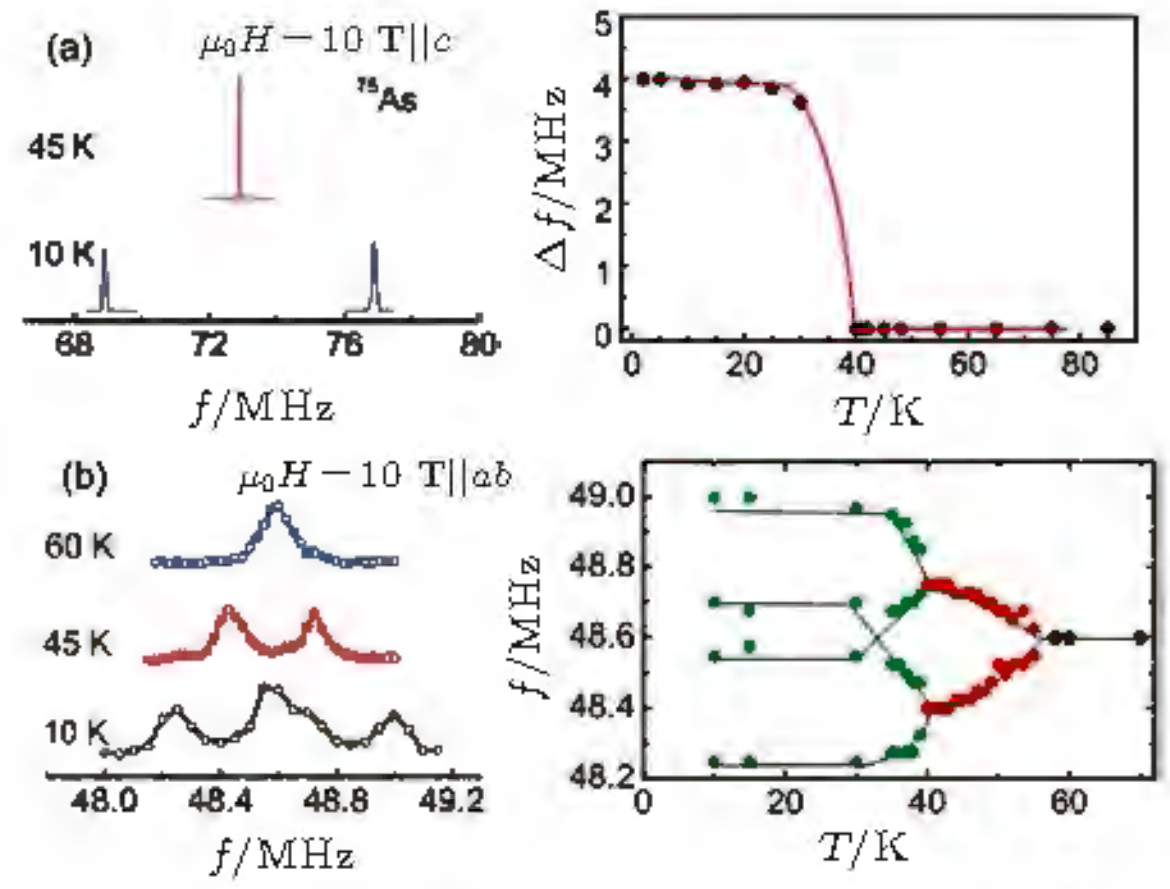

**Fig. 9.** (a) The line splitting of the $^{75}$As center transition in NaFeAs due to the internal field along the $c$ axis emerged in the AFM state. The inset shows line shift as the order parameter versus temperature. (b) The line splitting of the $^{75}$As satellite illustrating the structural transition. The inset shows the $^{75}$As satellite frequencies at different temperatures. The second splitting with decreasing temperature is due to the slight misalignment of the field to the crystalline $ab$ plane.[28]

The temperature dependences of $1/(^{75}T_1T)$ in NaFeAs single crystal are shown in Fig. 10(a) to study the nature of the structural transition. From 300 K to 100 K, $1/(^{75}T_1T)$ decreases with increasing temperature, which will be discussed later. Below 100 K, the $1/(^{75}T_1T)$ shows a mild upturn behavior. Besides, $1/(^{75}T_1T)$ is anisotropic with different field orientations. Both the behaviors indicate the development of strong low-energy spin fluctuations. The spin-lattice relaxation rates can be fitted by the Curie–Weiss behavior $1/(T_1T) = A/(T+\theta)$. With the field used in the crystalline *ab* plane, the values of fitting parameter $\Theta$ are determined to be $\theta \sim -10\pm5$ K (above $T_S$) and $\theta \sim -40\pm1$ K (below $T_S$), indicating that the spin fluctuations are strongly enhanced in the orthorhombic phase. This suggests that the structural transition and the magnetism are strongly coupled. However, as shown in Fig. 10(b), just below $T_S$, we do not see significant change of $1/T_1$ with the field rotating in the *ab* plane.

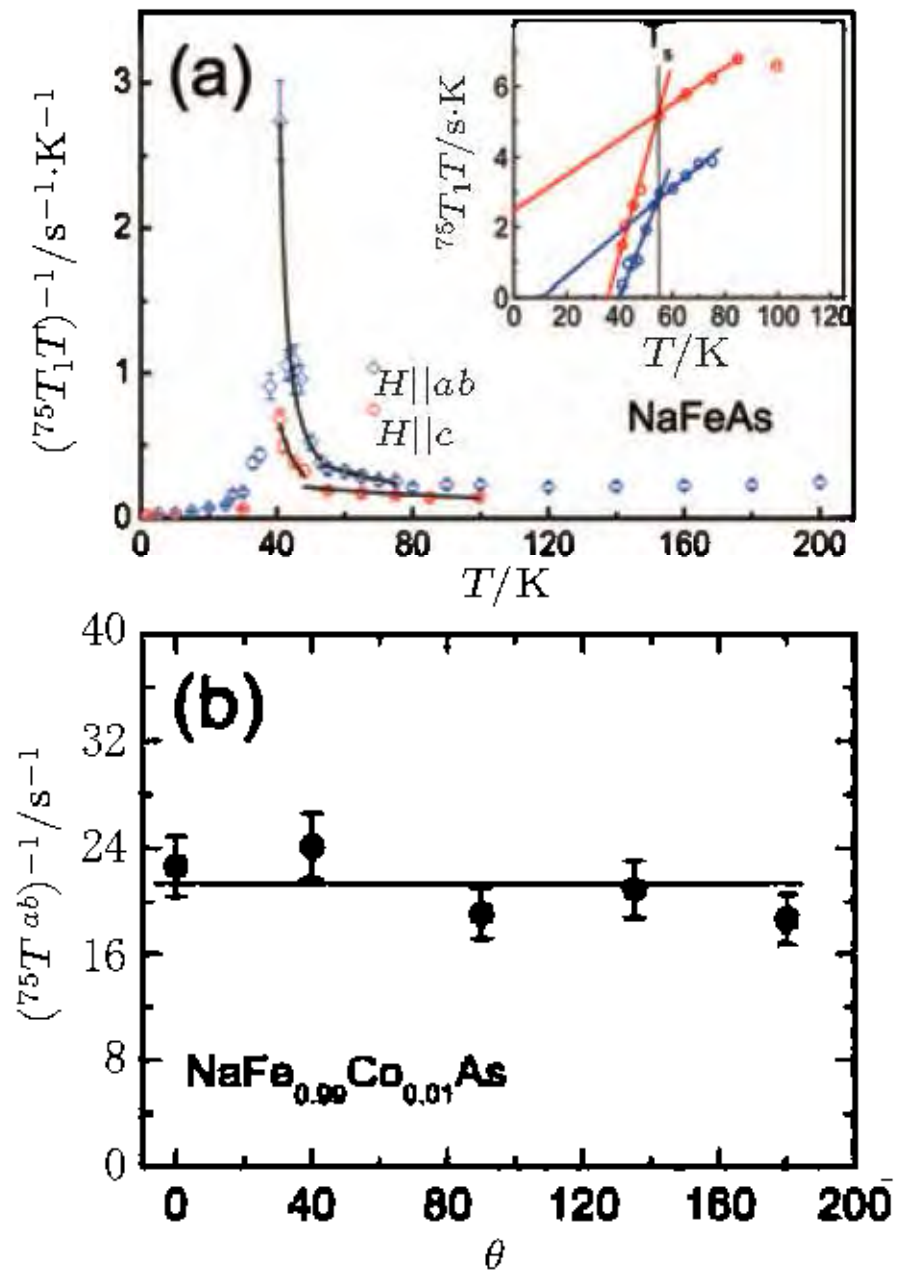

**Fig. 10.** (a) Plots of $1/(^{75}T_1T)$ versus temperature in NaFeAs single crystals with a field of 10 T. The inset shows $^{75}T_1T$ versus temperature. The solid lines are fittings to $^{75}T_1T \sim (T+\theta)$.[28] (b) The angular dependence of $1/T_1$ below the structural transition in $NaFe_{0.99}Co_{0.01}As$ single crystal.

## 5. NMR study of LROAFM in iron-based materials

As a local probe, NMR is very useful to identify the LROAFM in iron pnictides. Kitagawa *et al.*[27] pointed out the configuration for the hyperfine coupling between $^{75}$As nucleus and Fe moments in the stripe AFM order, as shown in Fig. 11. The internal hyperfine field is written as

$$\boldsymbol{H}_{\text{int}} = \sum_{i=1}^{4} \boldsymbol{A}_i \boldsymbol{m}_i, \tag{22}$$

where $\boldsymbol{A}_i$ and $\boldsymbol{m}_i$ denote the coupling tensor and magnetic moment of the $i$-th Fe, respectively. For the first Fe site Fe1, the hyperfine coupling tensor can be written as

$$\boldsymbol{A}_1 = \begin{pmatrix} A_{aa} & A_{ab} & A_{ac} \\ A_{ba} & A_{bb} & A_{bc} \\ A_{ca} & A_{cb} & A_{cc} \end{pmatrix}. \tag{23}$$

Then, based on the mirror symmetry consideration, the hyperfine coupling tensor of other Fe sites can be obtained as

$$\boldsymbol{A}_2 = \begin{pmatrix} A_{aa} & -A_{ab} & -A_{ac} \\ -A_{ba} & A_{bb} & A_{bc} \\ -A_{ca} & A_{cb} & A_{cc} \end{pmatrix}, \tag{24}$$

$$\boldsymbol{A}_3 = \begin{pmatrix} A_{aa} & -A_{ab} & A_{ac} \\ -A_{ba} & A_{bb} & -A_{bc} \\ A_{ca} & -A_{cb} & A_{cc} \end{pmatrix}, \tag{25}$$

$$\boldsymbol{A}_4 = \begin{pmatrix} A_{aa} & A_{ab} & -A_{ac} \\ A_{ba} & A_{bb} & -A_{bc} \\ -A_{ca} & -A_{cb} & A_{cc} \end{pmatrix}. \tag{26}$$

For the stripe antiferromagnetic order,

$$\boldsymbol{m}_1 = -\boldsymbol{m}_2 = \boldsymbol{m}_3 = -\boldsymbol{m}_4 \equiv \boldsymbol{m}. \tag{27}$$

As a result, the internal hyperfine field is

$$\begin{aligned} \boldsymbol{H}_{\text{int}} &= (\boldsymbol{A}_1 - \boldsymbol{A}_2 + \boldsymbol{A}_3 - \boldsymbol{A}_4)\boldsymbol{m} \\ &= \begin{pmatrix} 0 & 0 & 4A_{ac} \\ 0 & 0 & 0 \\ 4A_{ca} & 0 & 0 \end{pmatrix} \begin{pmatrix} m_a \\ m_b \\ m_c \end{pmatrix} \\ &= 4A_{ac} \begin{pmatrix} m_c \\ 0 \\ m_a \end{pmatrix}. \end{aligned} \tag{28}$$

Based on these calculations, the stripe AFM in iron pnicitides will result in an internal hyperfine field along the crystalline $c$ axis.

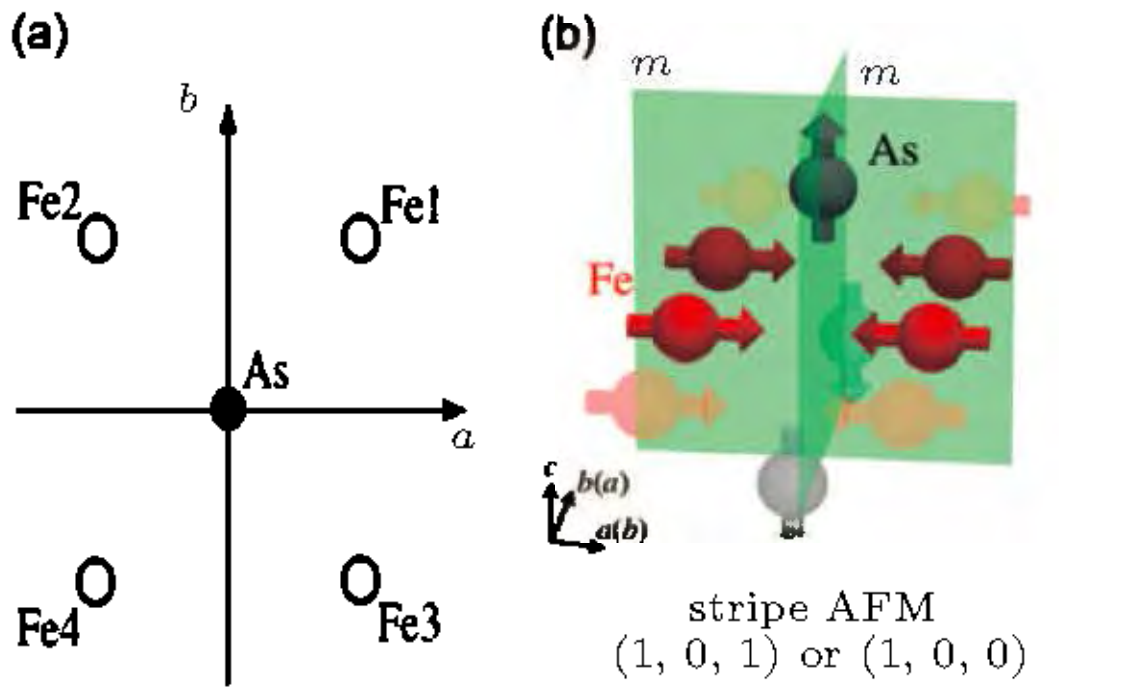

**Fig. 11.** Schematic configuration for the Fe–As tetrahedron in the orthorhombic unite cell as seen against the $c$ axis. (b) The stripe pattern LROAFM on Fe sites and the resultant internal hyperfine field on As sites.[27]

Figure 12 shows $^{75}$As NMR spectra across the antiferromagnetic transition in undoped $BaFe_2As_2$.[27] In the paramagnetic state, the spectrum is composed of a sharp central line and two satellites, as a result of the first-order quadrupole correction. Below the Neel temperature, the $c$-axis internal field splits the spectrum into peaks with $H||c$ and shifts them toward lower field with $H||ab$. From the analysis above, the field gap between the splitting central lines is proportional to Fe magnetic moment, which is determined to be 0.87 $\mu_B$ in $BaFe_2As_2$ from neutron scattering measurements.[29] As a result, the off-diagonal hyperfine coupling constant $A_{ac}$ is determined to be 0.43 $T/\mu_B$.

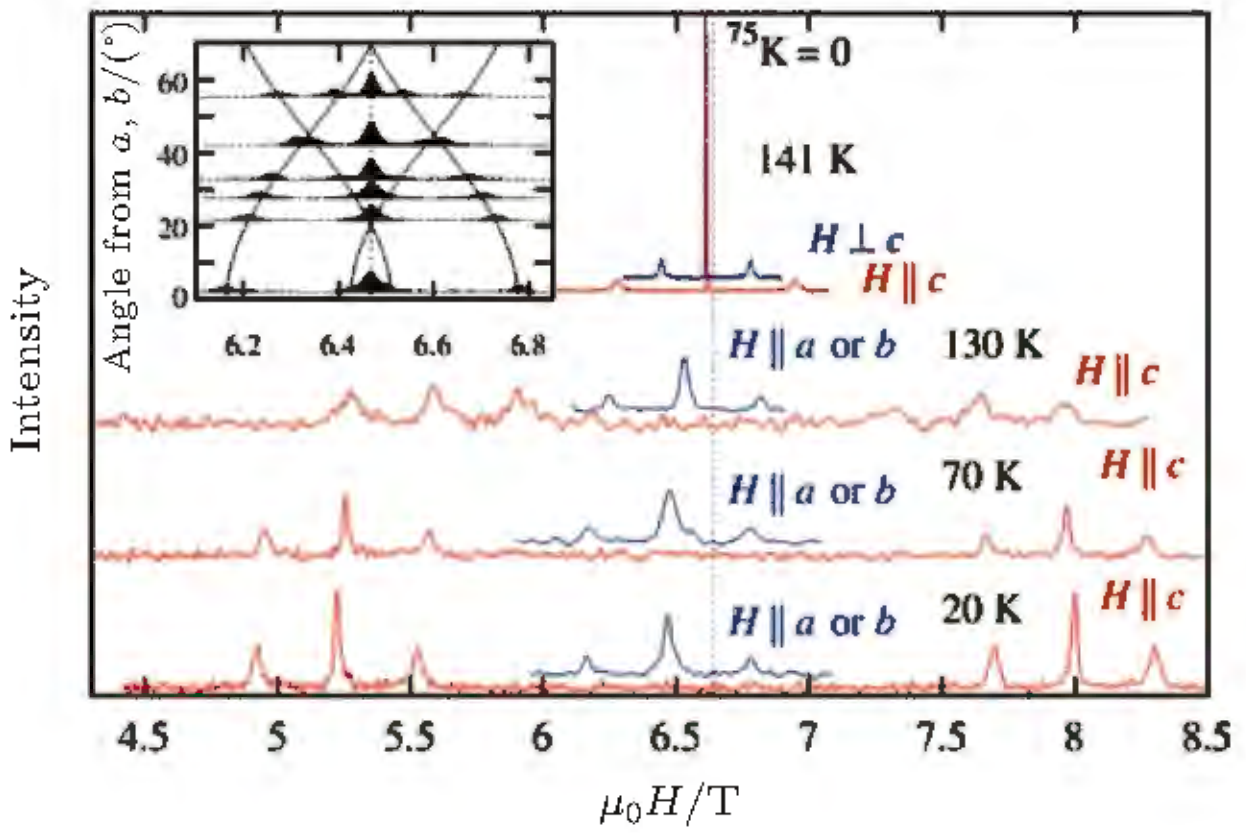

**Fig. 12.** A field-sweept $^{75}$As spectra (a single transition and two satellites) at 48.41 MHz for various temperatures in $BaFe_2As_2$ with different field orientations. The in-plane angular dependence of the spectra at $T$=20 K is shown in the inset.[27]

NaFeAs with the 111 structure has a separate structure and magnetic transition. The $^{23}$Na NMR spectra are reported by Ma *et al.*[28] to reveal the LROAFM, as shown in Fig. 13. The spectrum consists of a single central line and two satellites above $T_N$, and then each splits into two peaks below $T_N \sim 40.5$ K with $H||c$ due to the internal field along the crystalline $c$ axis. However, the line width and the magnetic moments show an unconventional behavior. First, the spectrum splitting $\Delta f$ of $^{23}$Na increases in two steps. The $\Delta f$ increases quickly below $T_N$, then increases slowly below 35 K and is saturated at $T = 10$ K. Second, the $^{23}$Na spectrum is clearly broadened between 40 and 30 K. With the temperature decreasing, the spectrum narrows again. For $^{75}$As, no signal is observed between 40 and 30 K, which indicates significant magnetic broadening of the spectrum. These behaviors are suggested to be induced by thermally activated domain walls, such as antiphase type or/and misaligned $a/b$ axis domains. At low temperatures, these domain walls are frozen. Another possible explanation is that the fluctuating features are caused by an incommensurate modulation on the AFM, which becomes commensurate again at low temperatures.[30] However, the $^{23}$Na spectrum between 40 and 30 K cannot be fitted by a conventional incommensurate magnetic order.

High-pressure NMR study of the antiferromagnetic order in NaFeAs is shown in Fig. 14. Determined by the paramagnetic spectral weight loss, the magnetic transtion temperature $T_N$ is raised by 18% at 2.56 GPa. The antiferromagnetic ordering moment is also raised by 30% at 2.56 GPa, which is observed from the line splitting of the $^{23}$Na NMR spectrum. Suppose that the lattice structure along the $c$ axis is more compressible than along $a$ or $b$ axis, the interlayer coupling $J_c$ may be very important for determining the antiferromagnetic transition and ordered moment based on the local moment picture.

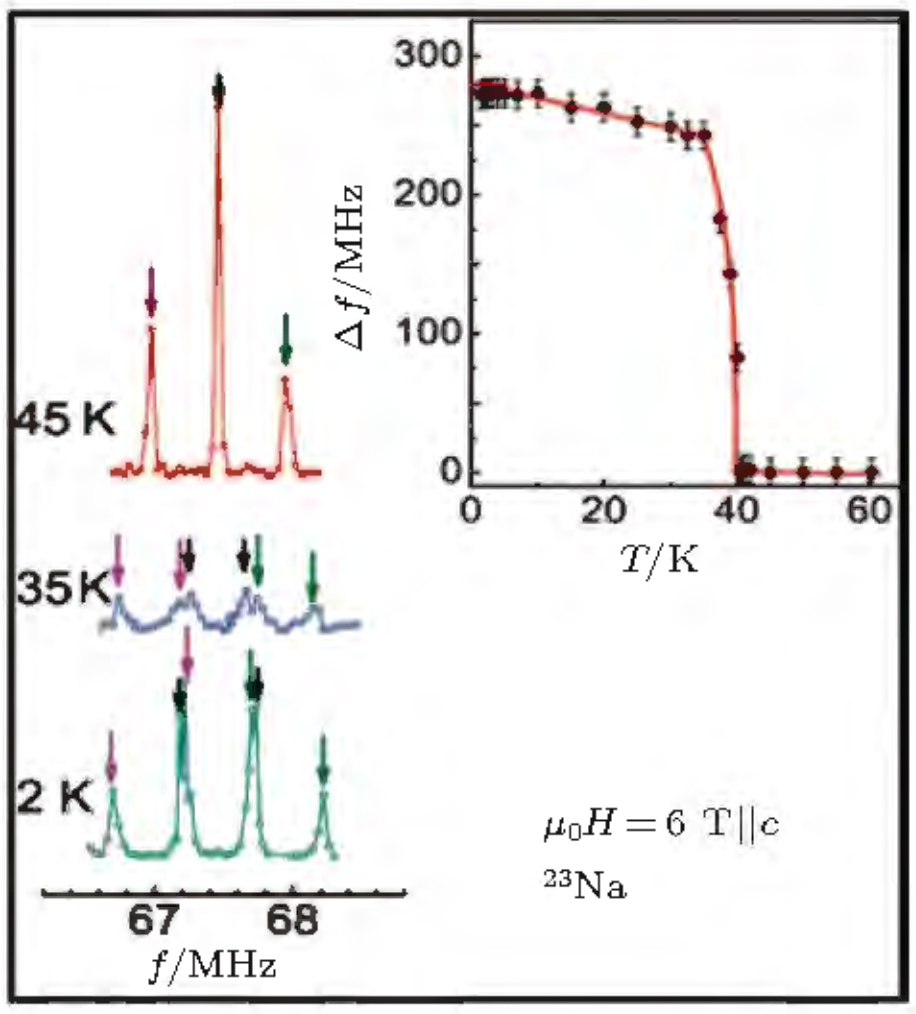

**Fig. 13.** The $^{23}$Na NMR spectra of NaFeAs single crystals at various temperatures with the field applied along the crystalline $c$ axis to show the antiferromagnetic transition. The inset shows the line shift of $^{23}$Na center transition as a function of temperature.[28]

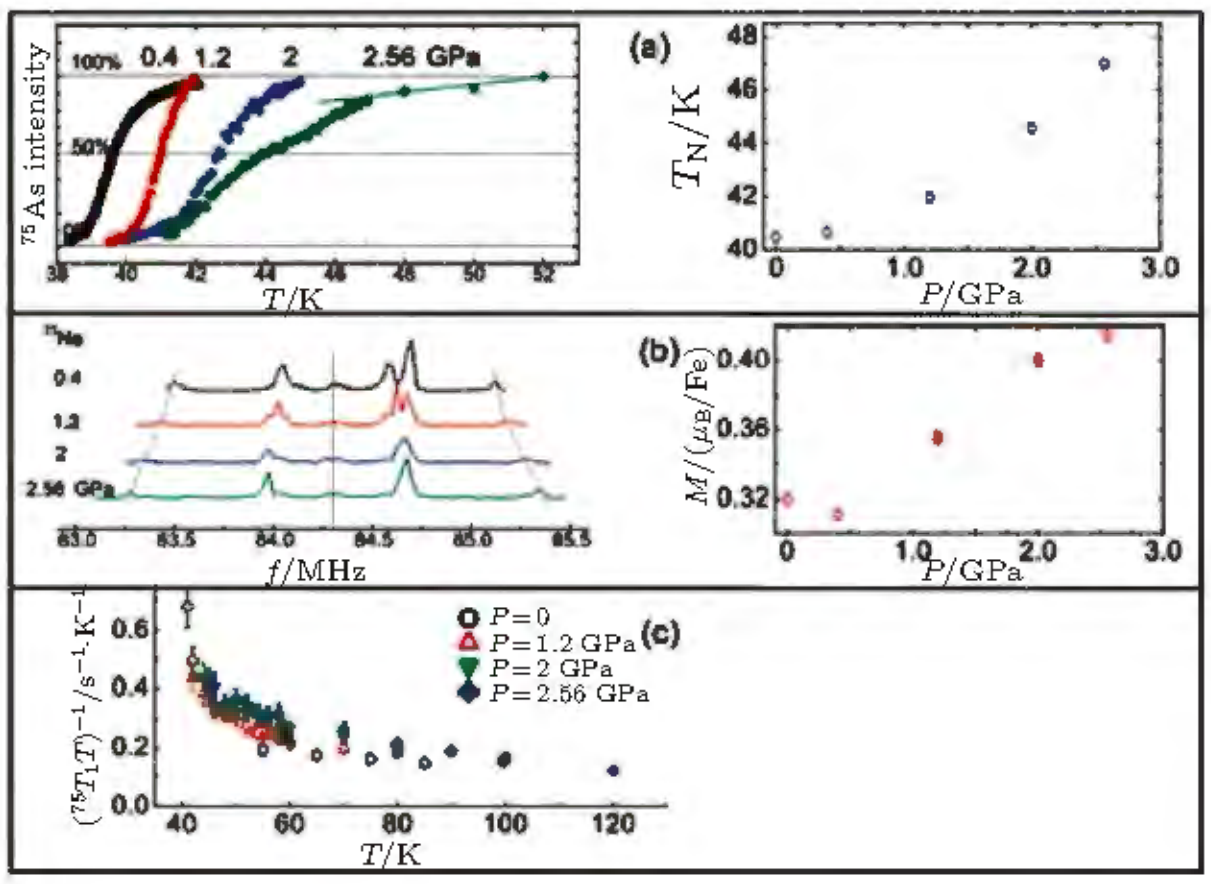

**Fig. 14.** (a) Plots of normalized $^{75}$As paramagnetic spectral weight versus temperature at various pressures of NaFeAs. The inset shows antiferromagnetic transition temperatures for different pressures. (b) $^{23}$Na spectra at $T = 2$ K, with the field applied along the $c$ axis at different pressures. The inset shows the pressure dependence of the antiferromagnetic moment. (c) The temperature dependences of $1/(^{75}T_1T)$ at various pressures.[28]

## 6. Coexistence and competition between LROAFM and SC

As a local probe, NMR is very useful to determine whether antiferromagnetic order microscopically coexists with SC in iron pnicitides, which is important for the understanding of high-$T_C$ superconductivity in iron pnictides. The microscopic coexistence has been recently studied by NMR in underdoped $BaFe_2(As_{1-x}P_x)_2$, $Ba(Fe_{1-x}Ru_x)_2As_2$, and $Ba_{1-x}K_xFe_2As_2$, which are reviewed below.

As shown in Fig. 15, Iye *et al.*[31] reported on their NMR evidence for the microscopic coexistence of LROAFM and SC in power of underdoped $BaFe_2(As_{1-x}P_x)_2$. In Fig. 15(a), we show the temperature dependence of $^{31}$P-NMR spectrum. Above $T_N$, a single sharp spectrum is observed in the powdered sample without quadrupole broadening for $^{31}$P nuclei ($I$=1/2). Below $T_N$, a broad NMR spectrum with gaussian shape gradually develops and coexists with a sharp peak. The broadened spectrum indicates the magnetic broadening in the underdoped sample, whose magnetic moments are summarized in Fig. 15(b). The magnetic moment and the internal hyperfine field on $^{31}$P sites increase steeply below $T_N$, and are strongly suppressed below $T_C^*$. As shown in Fig. 15(c), $(T_1T)^{-1}$ measured at the magnetic broad spectrum decreases at $T_C^*$, indicating the opening of superconducting gap in the magnetic region of the sample. The suppressed moment in the magnetic region below the superconducting transition is a direct evidence for the competition between these two orders. As shown in Figs. 15(d) and 15(e), the values of $(T_1T)^{-1}$ are measured at different frequencies across the broad spectrum at $T$=5 and 20 K, showing that the values of $(T_1T)^{-1}$ in all regions of the spectrum at 5 K are smaller than at 20 K, which confirms the SC in the entire region of the sample. Measurements by Iye *et al.* indicate the microscopic coexistence of and strong competition between SC and AFM in underdoped $BaFe_2(As_{1-x}P_x)_2$, suggesting that these two orders originate from the same Fermi surface.

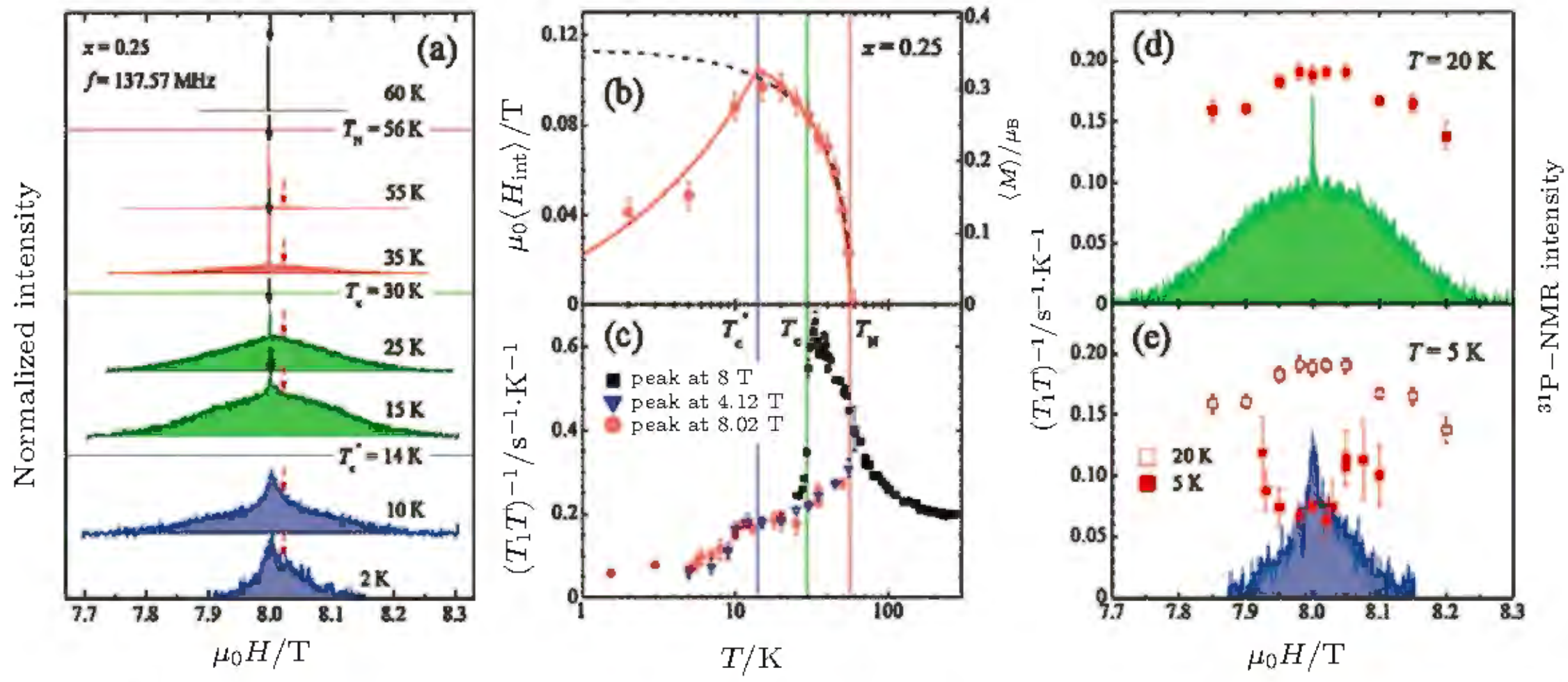

**Fig. 15.** (a) $^{31}$P NMR spectra at various temperatures in $BaFe_2(As_{0.75}P_{0.25})_2$ obtained by sweeping the magnetic field at the fixed frequency. (b) Plots of averaged hyperfine field on $^{31}$P sites $\langle H_{int}\rangle$ versus temperature, as obtained by the second moment of the spectrum. (c) Plots of $(T_1T)^{-1}$ measured at different positions of the spectra (solid squares are for the paramagnetic phase and the solid triangles and circles are for the antiferromagnetic phase) versus temperaure. Plots of $(T_1T)^{-1}$ across the spectra (d) above and (e) below the $T_C$ in the magnetically ordered phase.[31]

Broad doping range of coexistence between AFM and SC in the phase diagram is also suggested from the bulk measurements in the underdoped $Ba(Fe_{1-x}Ru_x)_2As_2$, which is an ideal system for characterizing whether the coexistence is microscopic or not.[32] As shown in Fig. 16, Ma *et al.*[33] reported on the $^{75}$As NMR spectra in the underdoped $Ba(Fe_{0.77}Ru_{0.23})_2As_2$ high-quality single crystals. Above $T_N \sim 60$ K, a single sharp central peak is observed. When temperature is lower than $T_N$, the spectrum begins to broaden and split into two peaks with equal frequency gap with $H||c$, and shifts to a higher frequency with $H||ab$, indicating a commensurate antiferromagnetic order. No evidence for residual paramagnetic phase is seen below $T_N$. However, full demagnetization is shown from the dc susceptibility measurements just below $T_C \sim 15$ K.

The spin dynamics shown by $1/(^{75}T_1T)$ in Fig. 17 also confirms this coexistence. As shown in Fig. 17(a), from 300 K to $T_N$, $1/(^{75}T_1T)$ is dominated by a Curie–Weiss upturn, indicating strong low-energy spin fluctuations. The antiferromagnetic transition is shown by the suppressed spin fluctuations below $T_N$ . Compared with the parent compound, $1/(^{75}T_1T)$ in $Ba(Fe_{1-x}Ru_x)_2As_2$ below $T_N$ tends to level off to a constant of one order higher before SC sets in, suggesting higher electron density of states on the Fermi surface by Ru dopants. On further cooling below 12 K, $1/(^{75}T_1T)$ drops abruptly, indicating the opening of superconducting gap in the antiferromagnetic region of the sample. Figure 17(b) shows the frequency dependences of $1/(^{75}T_1T)$ at $T = 20$ and 2 K. The reduced uniform value of $1/(^{75}T_1T)$ at 2 K supports the opening of a uniform superconducting gap below $T_C$. As shown in Fig. 17(c), the nuclear-spin relaxation curve shows clear single $T_1$ component behavior, indicative of the absence of phase separation. In Fig. 17(d), the values of $1/(T_1T)$s measured at different frequencies all drop obviously, indicating that the homogeneous superconducting gap is opened in the magnetically ordered region and leaves a rather large $1/(T_1T)$ at $T = 2$ K even when the high field is considered.

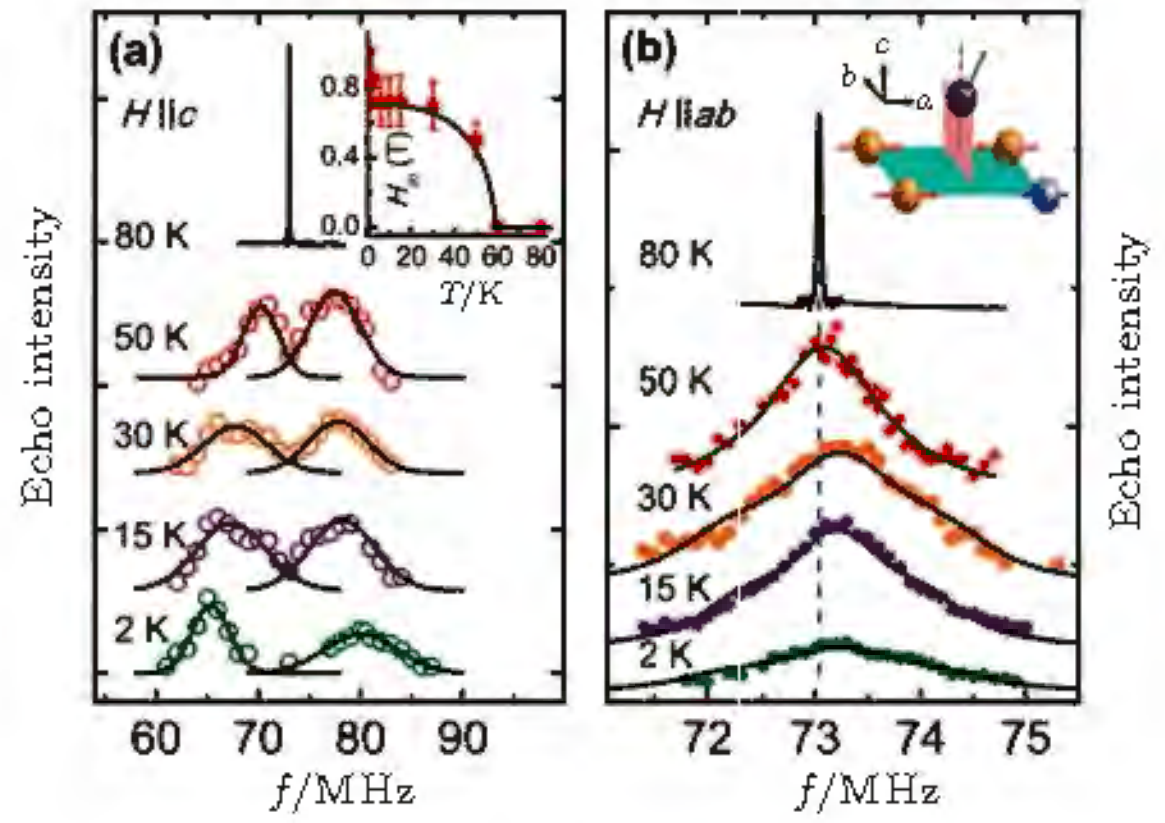

**Fig. 16.** $^{75}$As NMR spectra in $Ba(Fe_{0.77}Ru_{0.23})_2As_2$ single crystals at various temperatures with a 10-T field applied (a) along the $c$ axis and (b) in the $ab$ plane. The inset of panel (a) shows the temperature dependence of internal hyperfine field on $^{75}$As sites. The inset of panel (b) shows a schematic configuration of the Fe–As tetrahedron with one Ru substitution and the resultant hyperfine field on $^{75}$As sites.[32]

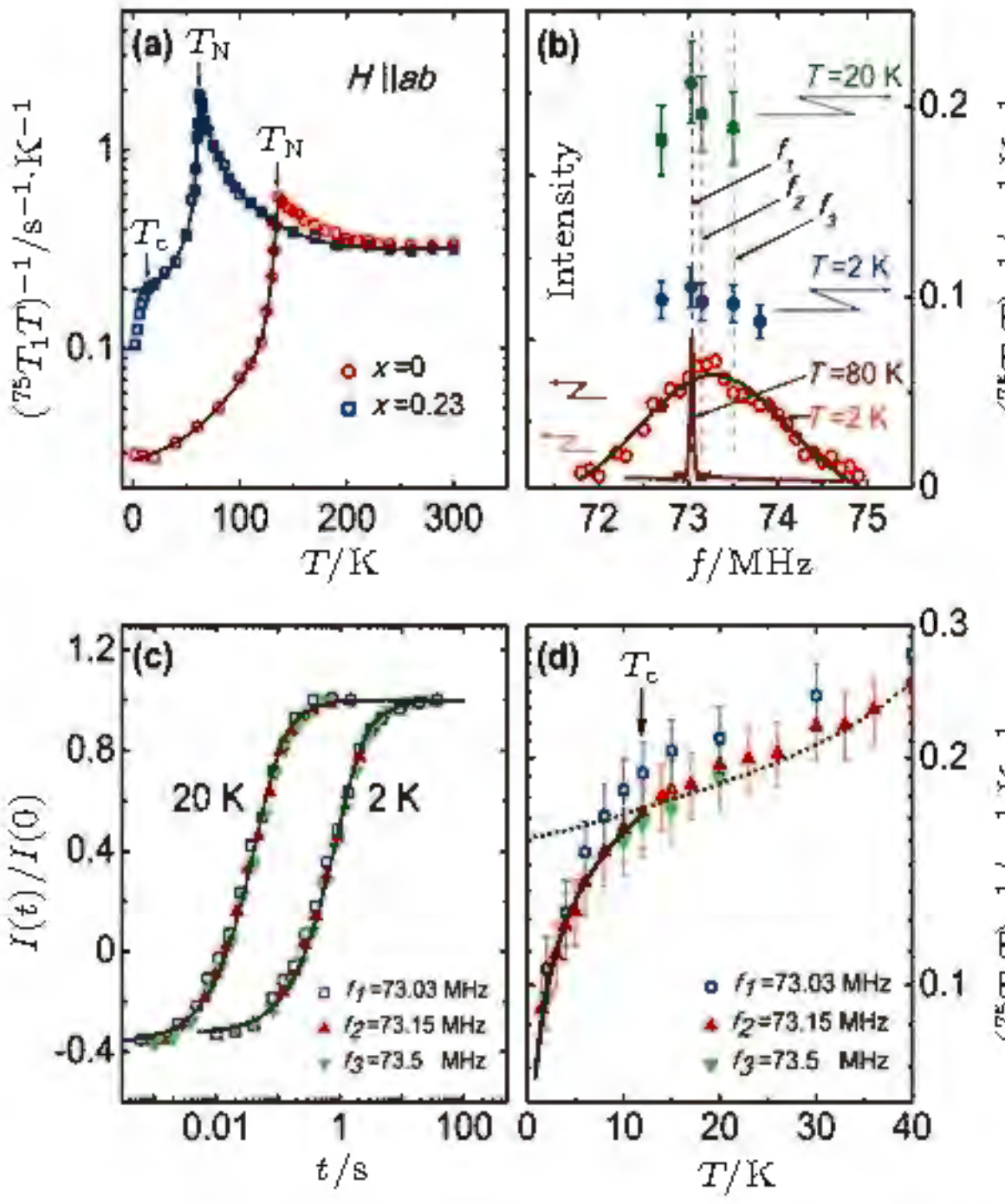

**Fig. 17.** (a) Temperature dependence of $1/(^{75}T_1T)$ in $Ba(Fe_{1-x}Ru_x)_2As_2$ ($x$=0, 0.23) with $H||ab$. (b) The $1/(^{75}T_1T)$ at $T$= 20 and 2 K across the spectra, shown with the spectra obtained at $T$ = 80 K and 2 K. (c) $^{75}$As nuclear-spin relaxation curves for 20 K and 2 K, showing clear single $T_1$ component behavior. (d) Plots of $1/(^{75}T_1T)$ versus $T$ in the low-temperature region for different frequencies.[32]

These observations confirm the Ru dopants induced superconducting itinerate electrons, the microscopic coexistence of AFM and a high density of low energy excitation below $T_C$ in underdoped $Ba(Fe_{1-x}Ru_x)_2As_2$. Different from what happens in $BaFe_2(As_{1-x}P_x)_2$, no evidence for the suppressed magnetic moment below $T_C$ is observed in $Ba(Fe_{1-x}Ru_x)_2As_2$. This is probably because the magnetic moment is large in our case.

The microscopic coexistence is also observed in underdoped $Ba_{1-x}K_xFe_2As_2$ single crystals, reported by Li *et al.*[34] As shown in Fig. 18, the AFM is monitored by the line splitting with $H||c$ and line shifting to a higher frequency with $H||a$ below $T_N \sim 46$ K. No evidence for residual paramagnetic phase is seen below $T_N$. Figure 19 displays the temperature dependence of spin-lattice relaxation rate measured at the central peak with $H||a$. Above $T_N$, the $1/T_1$ shows an upturn behavior with temperature decreasing due to strong spin fluctuations in a system near the magnetic instability. Below $T_N$, $1/T_1$ begins to drop for the gradually suppressed spin fluctuations in an ordered state. When the SC sets in below $T_C$, $1/T_1$ shows another sharp drop and follows a $T^3$ power law behavior, which is clear evidence for the microscopic coexistence of AFM and SC.

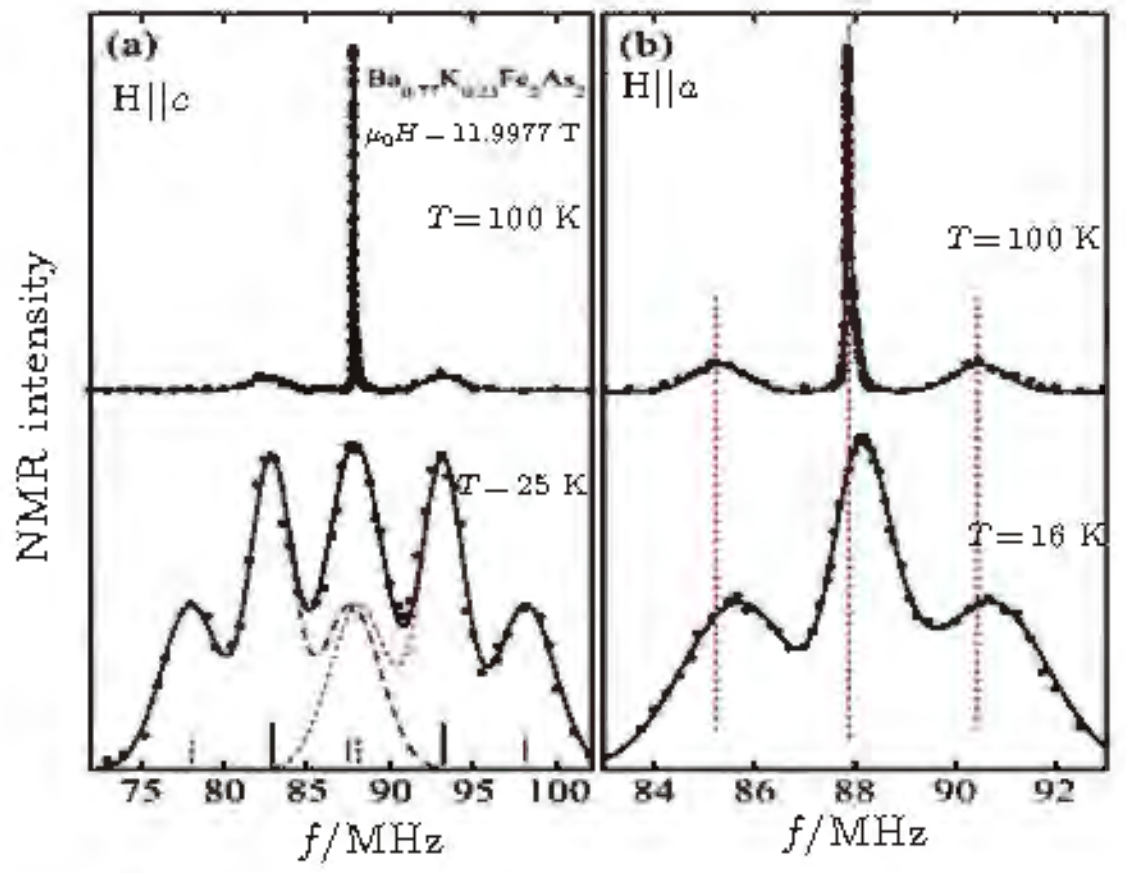

**Fig. 18.** $^{75}$As NMR spectra with the field applied (a) along the $c$ axis and (b) in the $ab$ plane of $Ba_{0.77}K_{0.23}Fe_2As_2$ single crystals above and below $T_N$.[34]

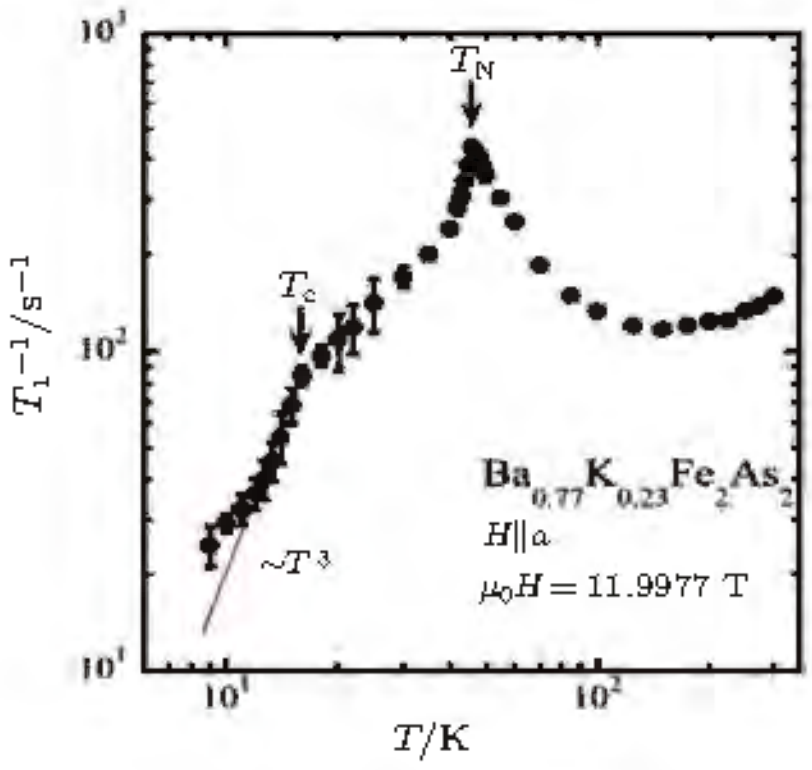

**Fig. 19.** $^{75}$As spin-lattice relaxation rate with $H||a$ on the logarithmic axes in $Ba_{0.77}K_{0.23}Fe_2As_2$. The solid line shows the power law behavior, $1/T_1 \propto T^3$.[34]

## 7. Normal state spin fluctuations

NMR, as a low energy probe, is very sensitive to low energy excitations in solids, which is of great importance for condensed matter physics. In iron pnictides, it is still controversial whether the magnetic order originates from an itinerate or a local picture. In this section, we will review the NMR study of the spin fluctuation properties in iron pnictides. Evidence for two types of spin fluctuations is found by NMR. We first review the formulation of the $1/(T_1T)$ in pnictides, and then review the studies on electron-doped $LaFeAsO_{1-x}F_x$, $Ba(Fe_{1-x}Co_x)_2As_2$, $LiFe_{1-x}Co_xAs$, and $NaFe_{1-x}Co_xAs$, the isoelectron-doped $BaFe_2(As_{1-x}P_x)_2$, the hole-doped $Ba_{1-x}K_xFe_2As_2$, $Li_{1+x}FeAs$, $K_yFe_{2-x}Se_2$, $(Tl,Rb)_yFe_{2-x}Se_2$, and $BaFe_2As_2$.

### 7.1. Low-energy spin fluctuation and the quantum criticality

Kitagawa *et al.*[35] proposed that the anisotropic $1/(T_1T)$ gives the information about the momentum dependence of spin fluctuations. The spin-lattice relaxation rate $1/T_1$ is analyzed as the contribution from the hyperfine field fluctuations perpendicular to the applied field, and it is given by

$$\left(\frac{1}{T_1}\right)_z = \frac{(\mu_0\gamma_N)^2}{2}\int_{-\infty}^{\infty} dt\, e^{i\omega t}(\langle H_{hf,x}(t), H_{hf,x}(0)\rangle + \langle H_{hf,y}(t), H_{hf,y}(0)\rangle) = (\mu_0\gamma_N)^2(|H_{hf,x}(\omega)|^2 + |H_{hf,y}(\omega)|^2). \quad (29)$$

Based on the hyperfine coupling tensor shown in Section 5, the fluctuating hyperfine field $H_{hf}(\omega)$ is obtained with Fe spin fluctuations $S(\omega)$ as shown below, for uncorrelated spin fluctuations,

$$\begin{pmatrix}(1/T_1)_{H||a}\\(1/T_1)_{H||b}\\(1/T_1)_{H||c}\end{pmatrix} \propto \begin{pmatrix}|A_{bb}S_b(\omega)|^2+|A_{cc}S_c(\omega)|^2\\|A_{cc}S_c(\omega)|^2+|A_{aa}S_a(\omega)|^2\\|A_{aa}S_a(\omega)|^2+|A_{bb}S_b(\omega)|^2\end{pmatrix}, \quad (30)$$

for stripe ($\pi$, 0) correlations,

$$\begin{pmatrix}(1/T_1)_{H||a}\\(1/T_1)_{H||b}\\(1/T_1)_{H||c}\end{pmatrix} \propto \begin{pmatrix}|A_{ac}S_a(\omega)|^2\\|A_{ac}S_a(\omega)|^2+|A_{ac}S_c(\omega)|^2\\|A_{ac}S_c(\omega)|^2\end{pmatrix}, \quad (31)$$

for checkerboard ($\pi$, $\pi$) correlations,

$$\begin{pmatrix}(1/T_1)_{H||a}\\(1/T_1)_{H||b}\\(1/T_1)_{H||c}\end{pmatrix} \propto \begin{pmatrix}|A_{ab}S_a(\omega)|^2\\|A_{ab}S_b(\omega)|^2\\|A_{ab}S_a(\omega)|^2+|A_{ab}S_b(\omega)|^2\end{pmatrix}. \quad (32)$$

Suppose that the Fe spin fluctuation is isotropic, the anisotropy of $1/T_1$ defined as $R \equiv (1/T_1)_{H||ab}/(1/T_1)_{H||c}$ is obtained, where the uncorrelated fluctuation gives

$$R = \frac{(A_{bb}^2+A_{cc}^2+A_{cc}^2+A_{aa}^2)/2}{A_{aa}^2+A_{bb}^2} = \frac{A_{aa}^2+A_{cc}^2}{2A_{aa}^2} = 0.5+0.5\left(\frac{A_{cc}}{A_{aa}}\right)^2, \quad (33)$$

stripe-pattern correlation gives

$$R = \frac{(A_{ac}^2+2A_{ac}^2)/2}{A_{ac}^2} = 1.5, \quad (34)$$

and checkerboard correlation gives

$$R = \frac{(A_{ab}^2+A_{ab}^2)/2}{2A_{ab}^2} = 0.5. \quad (35)$$

#### 7.1.1. Electron-doped pnictides

In Fig. 20, Kitagawa *et al.*[35] shows their NMR results about the $1/T_1$ anisotropy $R$ of $LaFeAsO_{1-x}F_x$ ($x = 0.07$, 0.11, and 0.14). For the samples with $x = 0.07$ and 0.11, $R$ is around 1.5 in the normal state and decreases below $T^*$, indicating the existence of strong stripe correlation in the normal state. However, in the $x = 0.14$ sample, $R$ is about 1.2 at $T = 225$ K and decreases to 1 at $T_C$, indicating an uncorrelated spin fluctuations in the overdoped samples. This evolution is reasonable since the stripe-like magnetic instability weakens with fluorine doping.

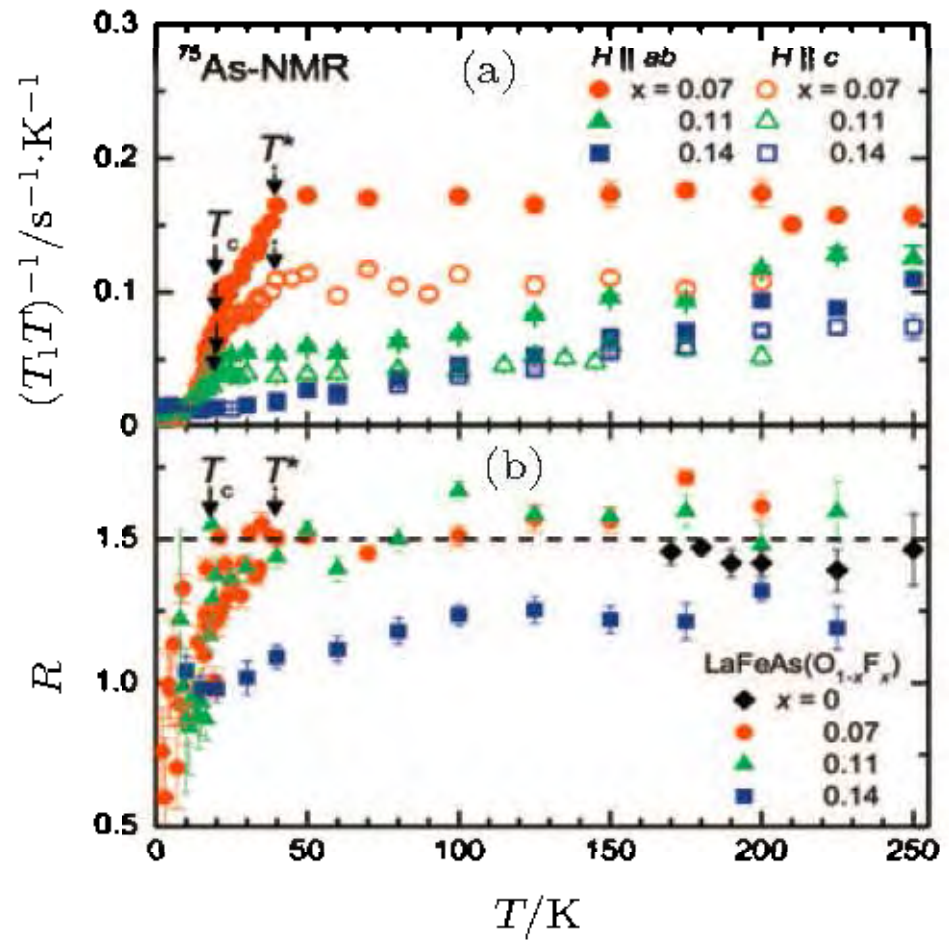

**Fig. 20.** (a) Plots of $^{75}$As $1/(T_1T)$ versus temperature for $H||ab$ and $H||c$ in $LaFeAsO_{1-x}F_x$ with various values of $x$. (b) The temperature dependences of the anisotropy of $1/T_1$ ($R \equiv (1/T_1)_{H||ab}/(1/T_1)_{H||c}$).[35]

As shown in Fig. 21, Oka *et al.*[13] reported on the $1/(T_1T)$ versus $T$ in the electron-doped $LaFeAsO_{1-x}F_x$ with 1111 structure to investigate the normal state spin fluctuation property. In the $x = 0.03$ sample, $1/(T_1T)$ shows strong upturn behavior with decreasing temperature and diverges at $T_N$, indicating the happening of antiferromagnetic transition. With the increase in doping level, the low-temperature upturn behavior is gradually suppressed until it totally vanishes at $x = 0.1$. The low-temperature upturn is not a Fermi liquid behavior, where constant $1/(T_1T)$ is expected. According to the theory for a weakly antiferromagnetically correlated quasi-2D metal, the upturn behavior can be reproduced by $1/(T_1T) = C/(T+\theta) + (1/T_1T)_0$. The first term on the right-hand side demonstrates the contributions to $1/(T_1T)$ from antiferromagnetic spin fluctuations and the second term shows the contributions to $1/(T_1T)$ from the Fermi surface, which is independent of temperature. The $\theta$ value essentially measures the spin fluctuation strength in the electronic system, which is also plotted in Fig. 22.

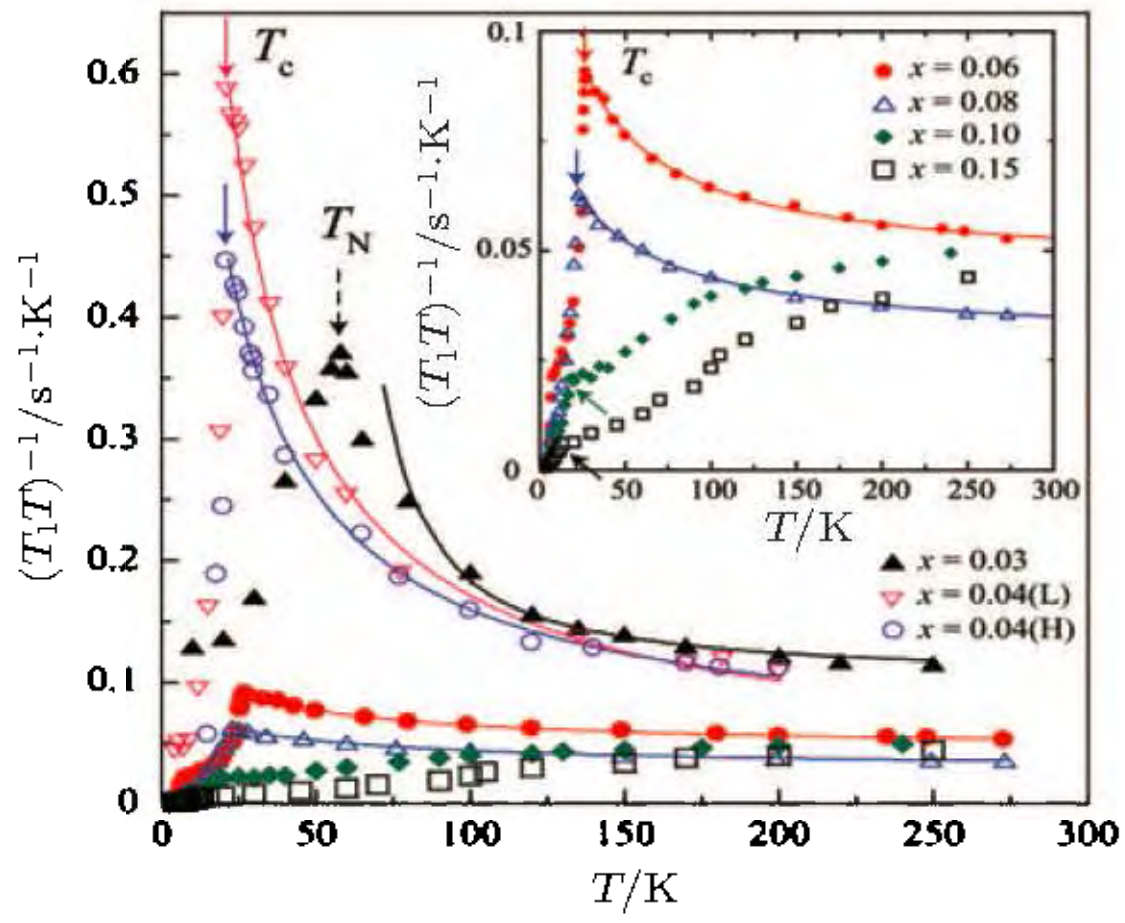

**Fig. 21.** Plots of $1/(T_1T)$ in $LaFeAsO_{1-x}F_x$ versus temperature. The solid lines represent the fittings to the theory for weakly antiferromagnetically correlated quasi-2D metal.[13]

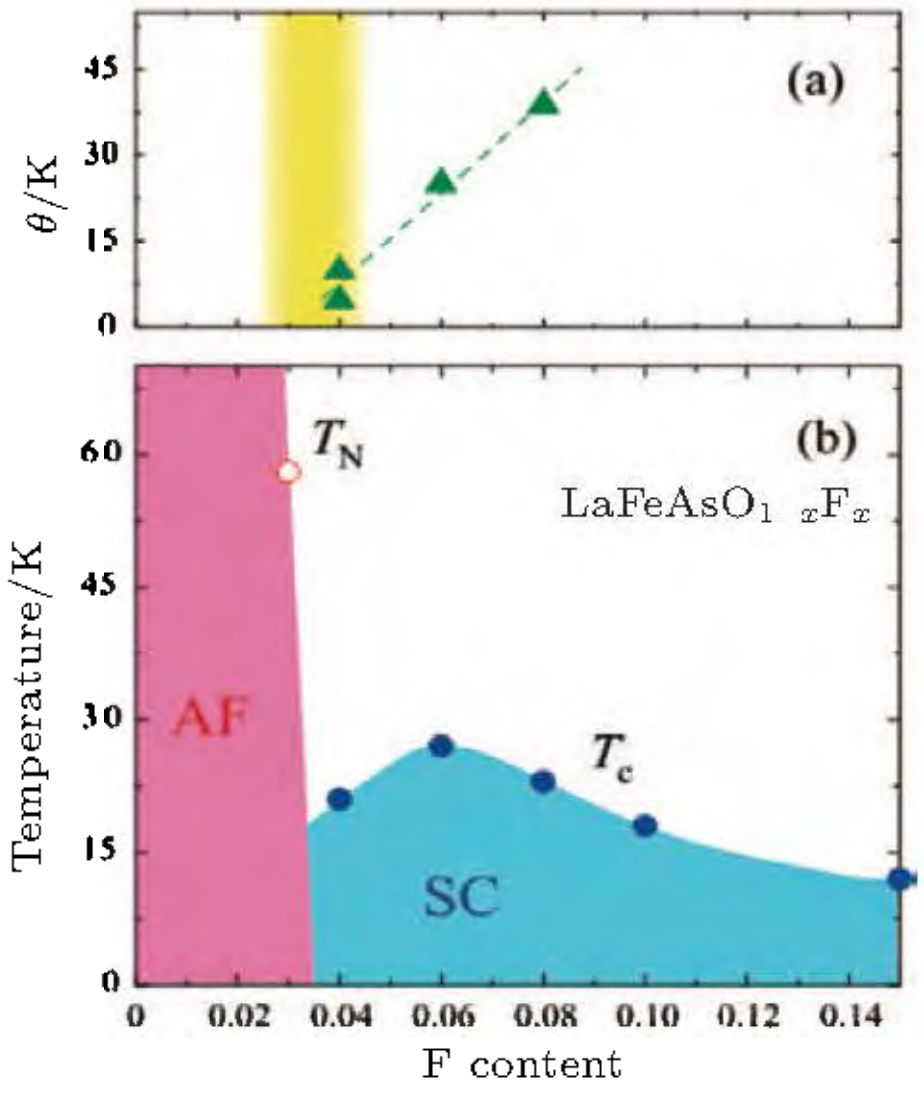

**Fig. 22.** (a) Doping dependence of the Weiss temperature $\theta$. (b) The phase diagram determined from this study.[13]

As illustrated above, the fitting parameter $\theta$ is plotted as a function of doping level in Fig. 22. With the doping level increasing, the low-energy spin fluctuations are gradually suppressed, as evidenced by increasing $\theta$ value. The existence of moderate spin fluctuation in the superconducting sample supports that spin fluctuation is a candidate pairing mechanism of superconductivity.

Ning *et al.*[36] reported on their NMR study on the evolutions of the spin fluctuations on electron-doped $Ba(Fe_{1-x}Co_x)_2As_2$ single crystals for various values of $x$, as shown in Fig. 23. The $1/(T_1T)$ data show an upturn characteristic upon cooling in samples with $x \leq 0.12$, indicative of the development of low-energy spin fluctuations. The decrease of the spin fluctuation strength is indicated from the suppressed low-temperature upturn behavior when the doping level increases from 0 to 0.12. Actually, the $1/(T_1T)$ data can be fitted by $1/(T_1T) = C/(T+\theta) + \alpha + \beta\exp(-\Delta/(k_BT))$, where the first term on the right-hand side refers to the low-energy spin fluctuations from interband scattering based on Moriya's theory for weakly antiferromagnetic itinerate system, and the second and third terms represent the phenomenologocal activation forms with unclear origin. The low-energy spin fluctuation strength can be quantitatively analyzed by the $\theta$ value or the value of $1/(T_1T)$ at $T = 25$ K.

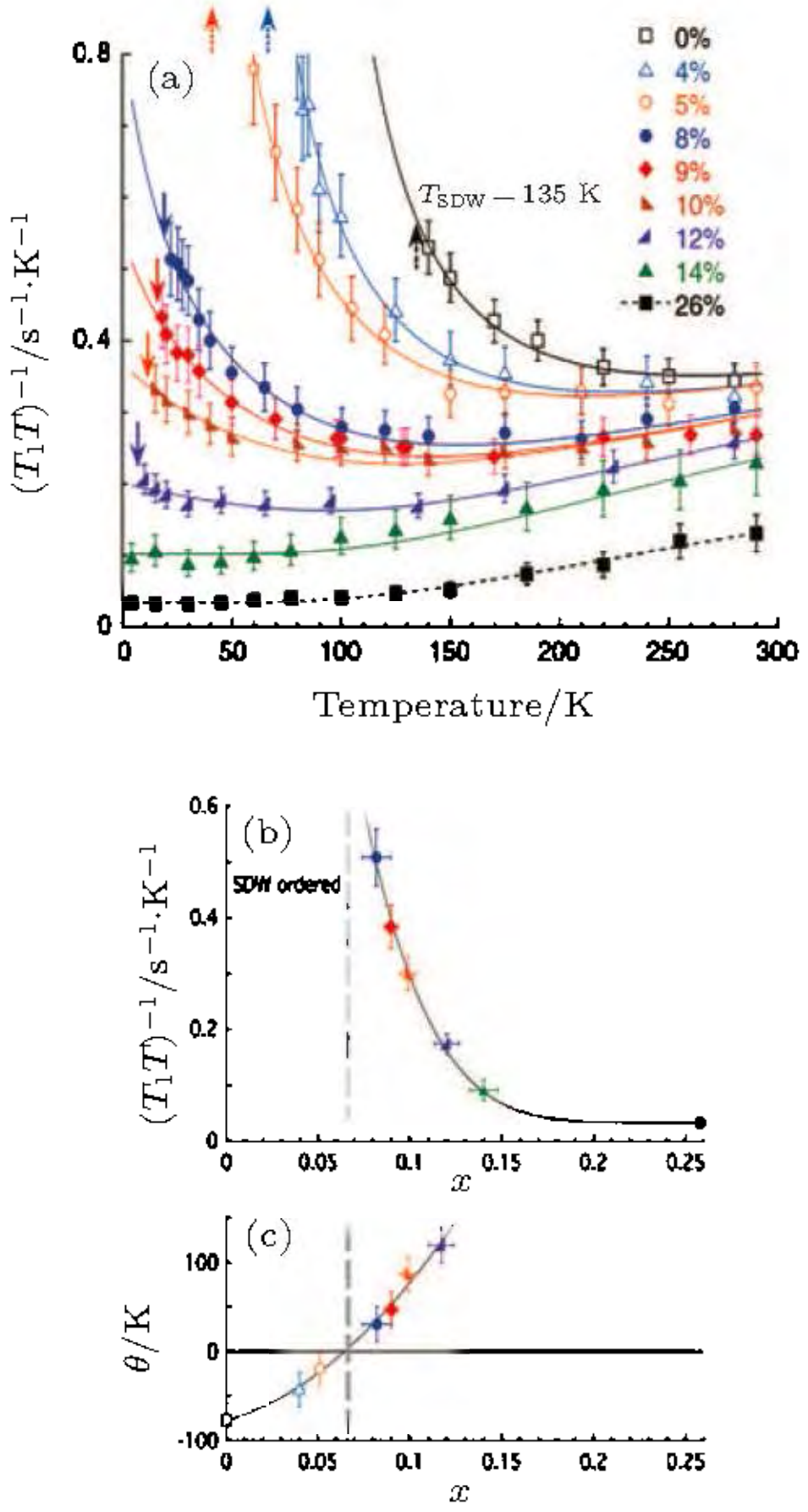

**Fig. 23.** (a) Plots of $^{75}$As $1/T_1T$ in $Ba(Fe_{1-x}Co_x)_2As_2$ versus temperature with an in-plane magnetic field, where $T_C$ and $T_N$ are denoted by the solid and dashed arrows. The solid and dashed lines represent the fittings to the function described in the main text, respectively, in order to analyze the spin fluctuation strength. The doping dependence of the low-energy spin fluctuation strength measured by $1/(T_1T)$ at $T = 25$ K and Weiss temperature $\theta$ are shown in panels (b) and (c), respectively.[36]

Figures 23(b) and 23(c) display the variations of $1/(T_1T)$ at 25 K and $\theta$ value with doping level. The $\theta$ value is negative for $x \leq 0.05$, indicative of the antiferromagnetic ground state for these samples, as shown in Fig. 23(c). In the meanwhile, the large value of $\theta \approx 119$ K for $x = 0.12$ overdoped sample indicates that it is far from the magnetic instability. The small positive $\theta \approx 31$ K for $x = 0.08$ optimally doped sample is an indication for the fact that the optimally doped superconductivity is slightly beyond the quantum critical point ($\theta = 0$). This work also suggests that the low-energy antiferromagnetic spin fluctuations may play an important role in high-$T_C$ superconductivity in the iron-based materials. Similar conclusions can also be drawn from FeSe under high pressure.[37]

### 7.1.2. Isoelectron-doped pnictides

The quantum critical behavior is slightly different in $BaFe_2(As_{1-x}P_x)_2$ as reported by Nakai *et al.*[38] As shown in Fig. 24, the low-energy spin fluctuation can be observed by the low-temperature upturn behavior. With the increase of doping level, this fluctuation is strongly suppressed and the Fermi liquid behavior emerges again in the overdoped $x$=0.64 sample. As shown in Fig. 25, the fitting parameter $\theta$ is also plotted as a function of the doping level, showing the importance of the low-energy spin fluctuation for the high-$T_C$ SC in iron pnictides. However, the optimally doped SC is located at the quantum critical point, which is different from the case of $LaFeAsO_{1-x}F_x$ and $Ba(Fe_{1-x}Co_x)_2As_2$.

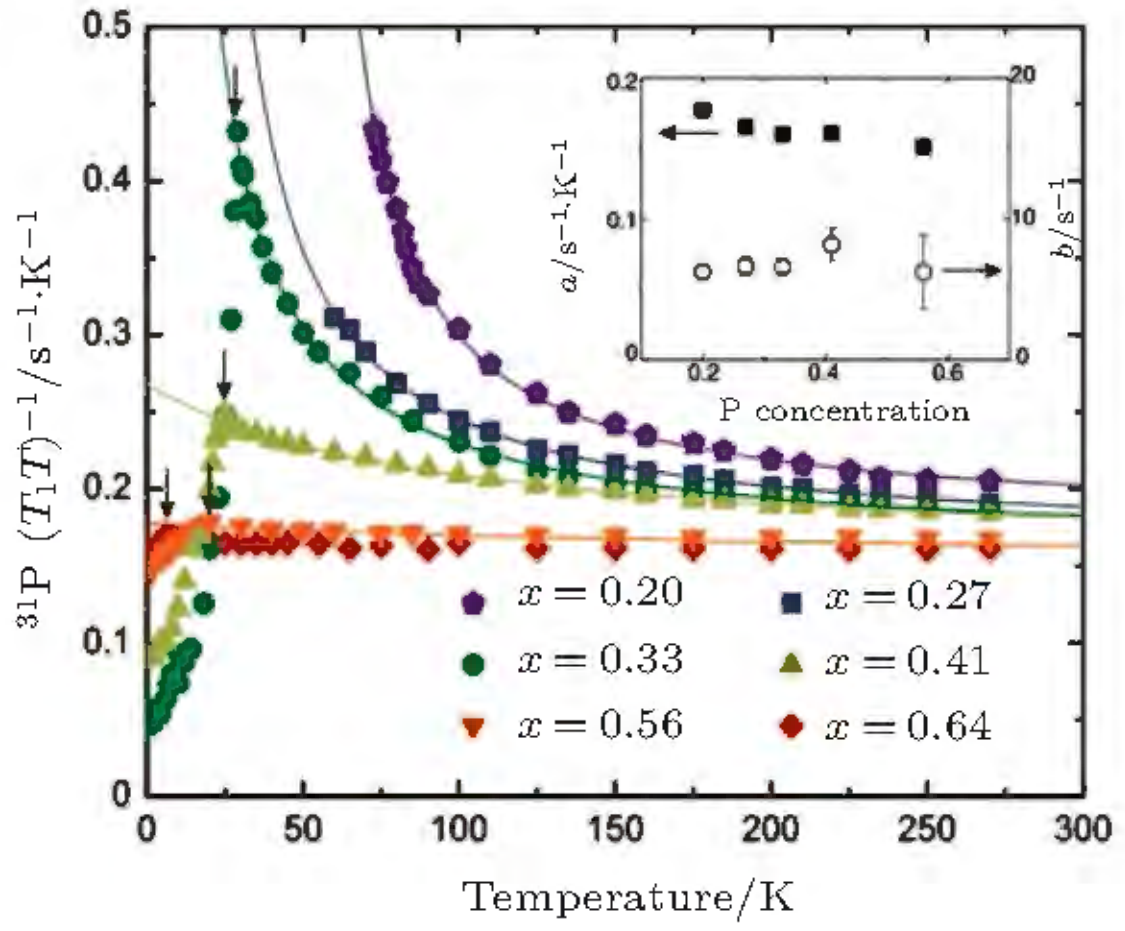

**Fig. 24.** Temperature dependence of $(T_1T)^{-1}$ in $BaFe_2(As_{1-x}P_x)_2$ with various values of $x$. Solid lines represent the fitting results to $(T_1T)^{-1} = a + b(T+\theta)^{-1}$, where the fitting parameters, $a$ and $b$, are shown in the inset.[38]

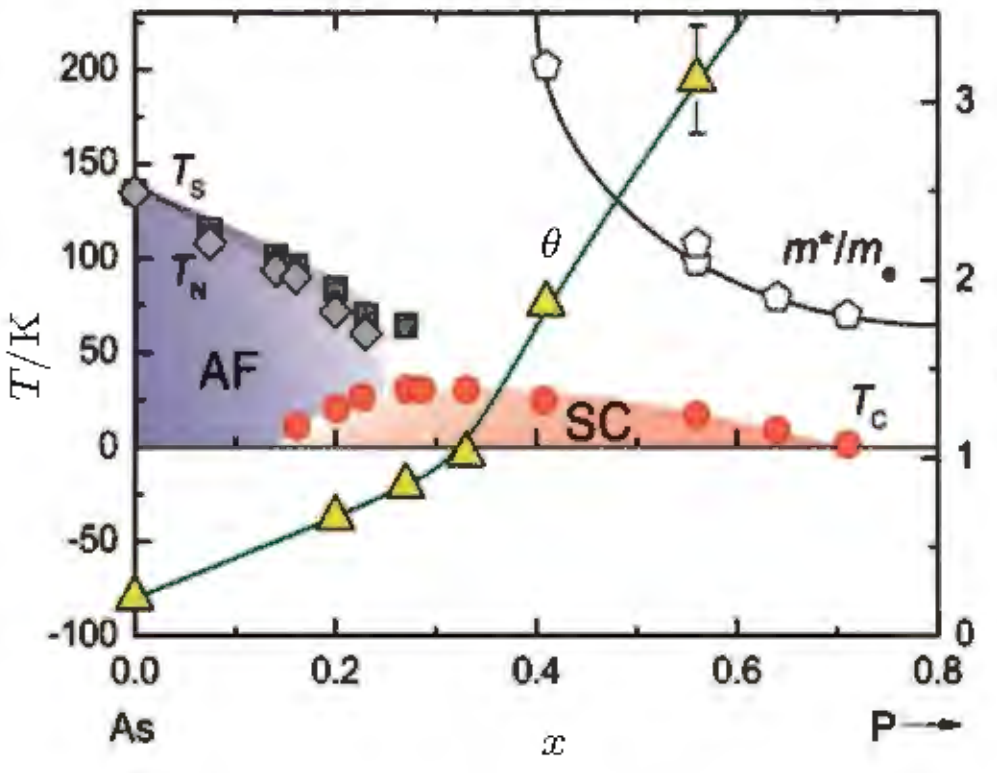

**Fig. 25.** Doping dependence of $\theta$ in $BaFe_2(As_{1-x}P_x)_2$ determined by NMR.[38]

### 7.1.3. Hole-doped pnictides

Being contradictory with the discussion above, strong low-energy spin fluctuation still exists in the heavily hole-doped $KFe_2As_2$ with 0.5 hole per Fe as reported by Zhang *et al.*[39] We replot the temperature dependences of $1/(T_1T)$ in $Ba_{1-x}K_xFe_2As_2$ for various values of $x$ with $H||ab$ in Fig. 26.[27,39–41] The divergence of $1/(T_1T)$ at $T_N$ ~135 K in $BaFe_2As_2$ indicates the antiferromagnetic transition. As shown in Fig. 26(a), with the doping level increasing from 0 to 0.7, the spin fluctuation strength gradually decreases as illustrated by the suppression of the upturn behavior, indicating the deviation from the magnetic instability. However, when $x$ increases up to 1, strong low-energy spin fluctuations still exist, which is shown in Fig. 26(b). This spin fluctuation is proved later to be stripe but incommensurate spin fluctuations by inelastic neutron scattering,[42] consistent with the anisotrpy of $1/T_1$ around 1.5 as shown in the inset of Fig. 26(b). This result is clearly different from the electron-doped case, where both the low-energy spin fluctuation and SC disappear quickly with the increae of doping content. Therefore, the data in $KFe_2As_2$ support a magnetic origin of the SC, and also prove that the more extended SC dome in the hole-doped case is due to the existence of low-energy spin fluctuations.

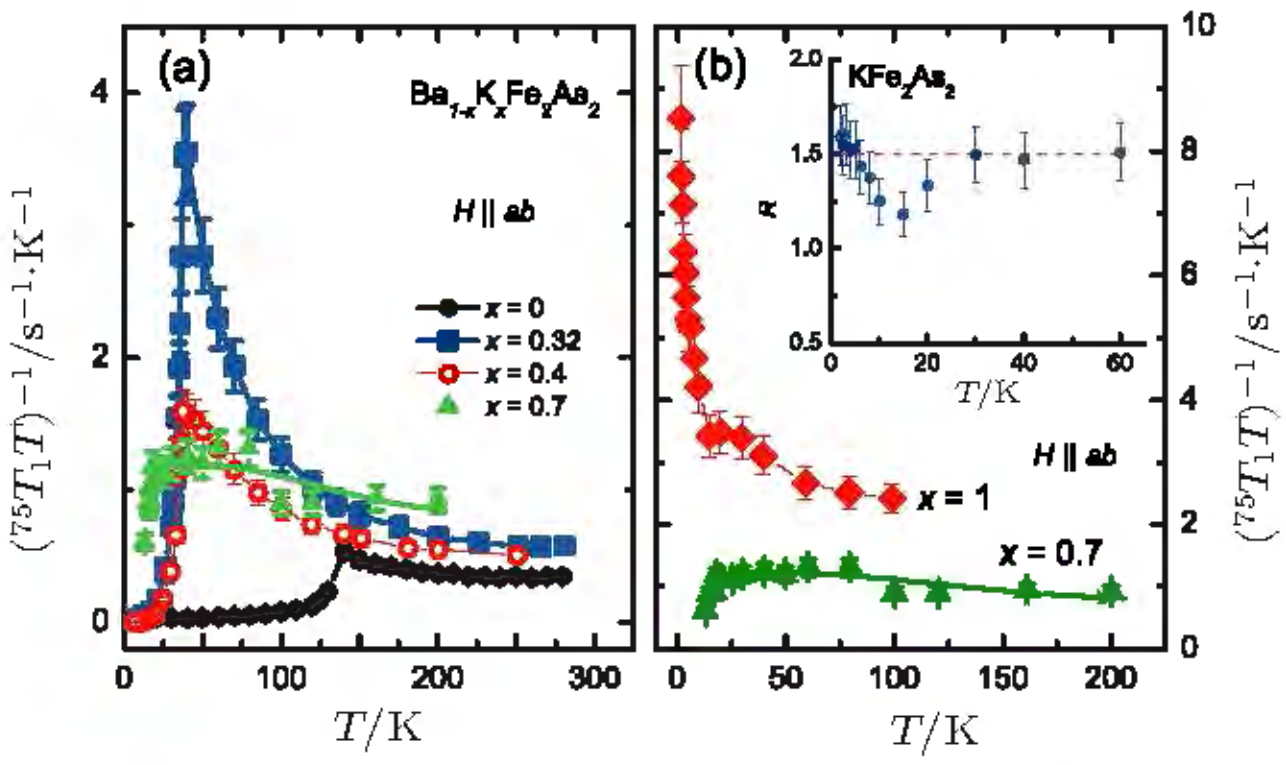

**Fig. 26.** (a) and (b) Temperature dependences of $1/(^{75}T_1T)$ in $Ba_{1-x}K_xFe_2As_2$ for various values of $x$ with field applied in the crystalline $ab$ plane. The inset in panel (b) shows the temperature dependence of the anisotropy of $1/T_1$ in $KFe_2As_2$.[27,39–41]

Strong low-energy spin fluctuation is also observed by Ma *et al.*[43] in $Ca_{1-x}Na_xFe_2As_2$ ($T_C$ ~ 32 K) with the doping level as high as 0.67. Figure 27 displays the temperature dependences of $1/(^{75}T_1T)$ with different field orientations in $Ca_{0.33}Na_{0.67}Fe_2As_2$. With the temperature decreasing, $1/(^{75}T_1T)$ shows strong upturn behavior with $H||c$, which can be fitted by the Curie–Weiss function, $1/(T_1T) = A/(T+\Theta)$. Additionally, the spin-lattice relaxation rate is anisotropic under different field orientations with an anisotropic factor $R \approx 1.3$. These observations supply strong evidence for strong low-energy spin fluctuations with stripe pattern in this system, although the doping level may be very high.

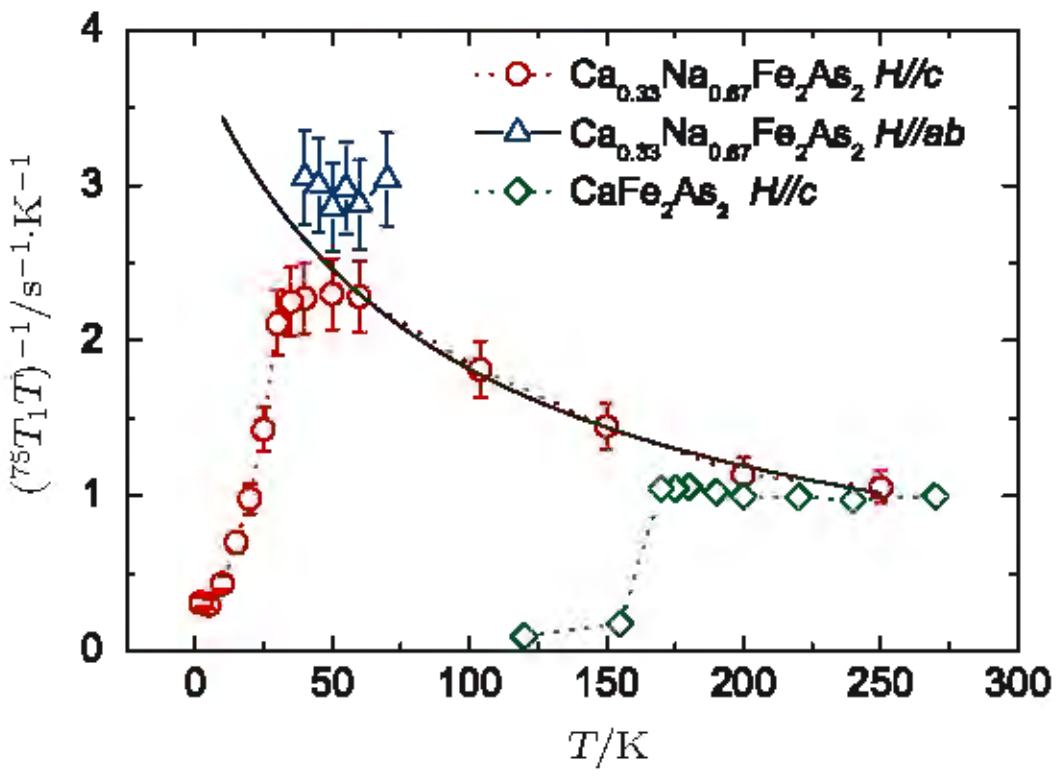

**Fig. 27.** The temperature dependences of $1/(^{75}T_1T)$ in $Ca_{0.33}Na_{0.67}Fe_2As_2$ single crystals with different field orientations. The solid line shows the fitting result to $1/(T_1T) = A/(T+\Theta)$. The $1/(^{75}T_1T)$ in $CaFe_2As_2$ is also shown as diamonds.[43,44]

#### 7.1.4. $Li_{1+x}FeAs$

LiFeAs has been proposed as an undoped superconductor. However, from NMR, Ma *et al.*[45] reported extra Li(2) sites, as shown in Figs. 28(a) and 28(b), which may explain the absence of spin fluctuations. Later experiments also suggest that it is a heavily electron-doped compound.[46] Figures 28(c) and 28(d) show the temperature dependences of $1/(T_1T)$ in $Li_{1+x}FeAs$.[45] With decreasing temperature, $1/(T_1T)$ shows strong upturn behavior under a field applied in the $ab$ plane. The $T_1$ anisotropy $(T_1^c/T_1^{ab})$ also increases as temperature drops, compared with that with $H||c$. These behaviors both indicate the development of stripe low-energy spin fluctuations and demonstrate that LiFeAs is also a strongly correlated system close to a magnetic ordering.

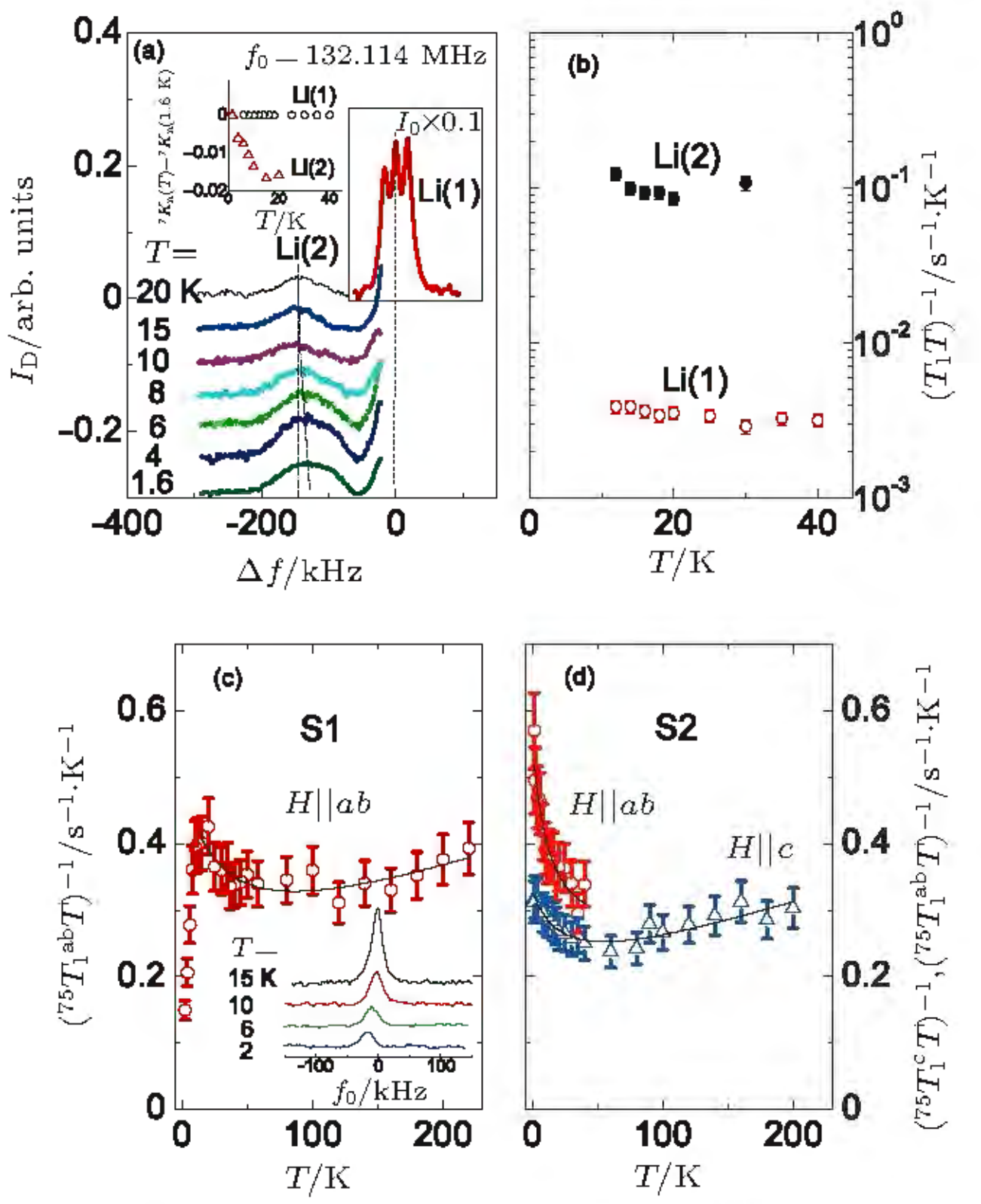

**Fig. 28.** (a) $^7$Li NMR spectra of a $Li_{1+x}FeAs$ crystal (S1) with an in-$ab$ plane 7.98-T field. The inset shows the normalized Knight shift of Li(1) and Li(2). (b) The spin-lattice relaxation rate measured at Li(1) and Li(2) sites. (c) Plots of $1/(^{75}T_1T)$ versus temperature $T$ of $Li_{1+x}FeAs$ sample S1 with an in-plane 8-T field. (d) Plots of $1/(^{75}T_1T)$ of sample S2 versus temperature $T$ with the field applied along the crystalline $c$ axis and in $ab$ plane. The solid lines in panels (c) and (d) represent the fitting results to the function, $1/(T_1T) = A/(T+\theta) + b + cT$ to study the low-energy spin fluctuations.[45]

#### 7.1.5. Itinerate nature of low-energy spin fluctuations

The low-energy spin fluctuations are usually observed in iron-based materials with the Fermi surface nesting effect. In order to figure out the complex relationship between the low-energy spin fluctuations and superconductivity, the orbital characteristics of this correlation are also studied, as the orbital degree of freedom should also play an important role in iron pnictides. For this purpose, Ye *et al.*[47] reported on the combined ARPES and NMR results of the $AFe_{1-x}Co_xAs$ ($A$ = Li, Na) system, as shown in Fig. 29. Increasing the doping level in $LiFe_{1-x}Co_xAs$, the Fermi surface nesting for the $d_{xy}$ hole pocket becomes better. Correspondingly, the low-temperature Curie–Weiss upturn in the $1/(T_1T)$ is also enhanced with the increase in Co doping content, as shown in Fig. 29(d), which is consistent with the evolution of the nesting effect. However, the SC in $LiFe_{1-x}Co_xAs$ is strongly suppressed by Co doping. On the contrary, in the optimally doped $NaFe_{0.955}Co_{0.045}As$, the nesting between $d_{xz}/d_{yz}$ hole pockets and electron pockets is nearly perfect. When more electrons are doped, both the $T_C$ and Fermi surface nesting are suppressed. These data suggest that the low-energy spin fluctuations are correlated with Fermi surface nesting from itinerate electrons. However, there is an orbital selective relation between the nesting and the superconductivity, where the nesting between the $d_{xz}/d_{yz}$ pockets, rather than the $d_{xy}$ pockets, is important for the high-temperature superconductivity in iron pnictides.

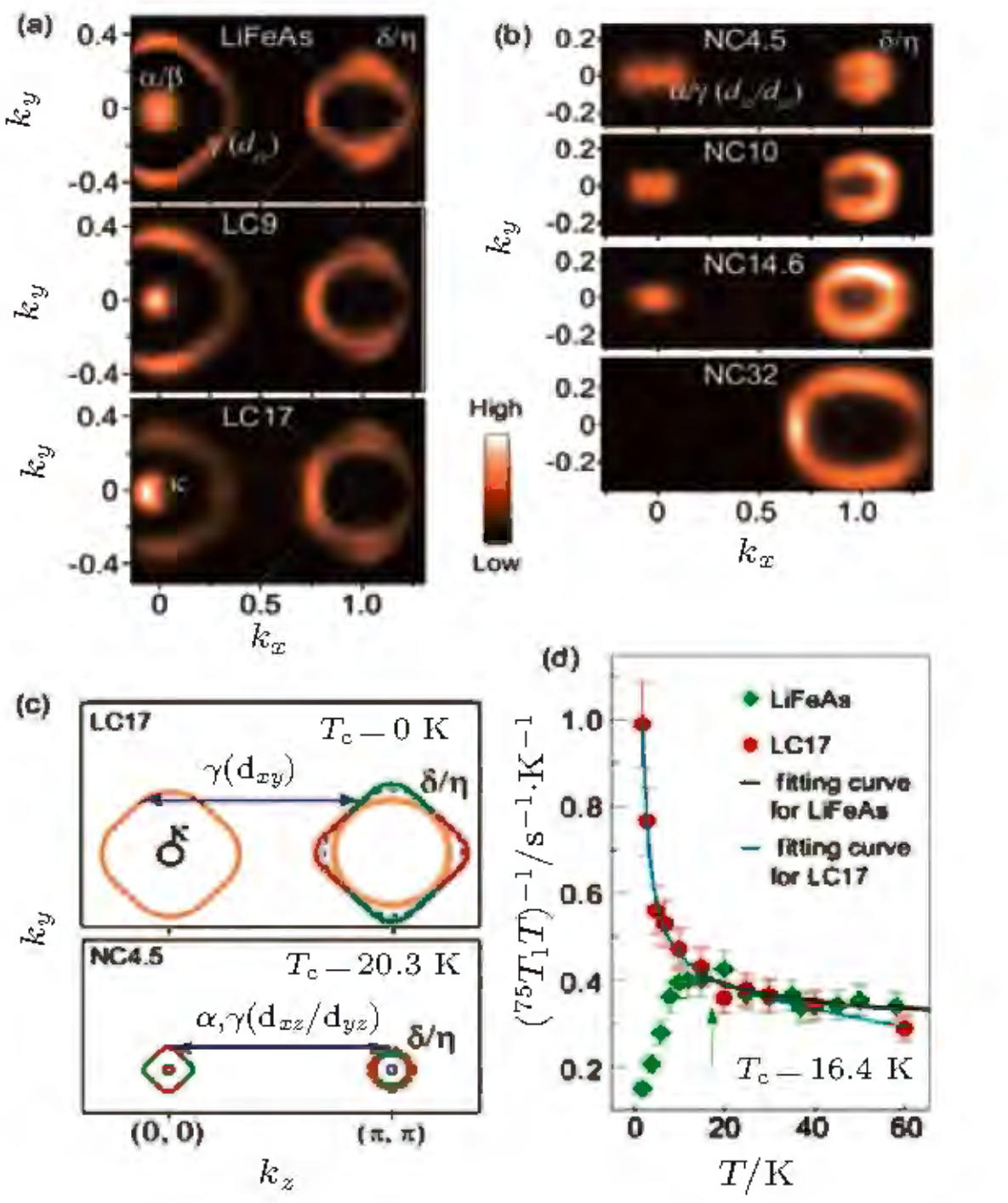

**Fig. 29.** (a) Fermi surfaces of $LiFe_{1-x}Co_xAs$ single crystals for various values of $x$, measured with 21.2 eV photons in mixed polarization. (b) The Fermi surfaces of $NaFe_{1-x}Co_xAs$ single crystals for various values of $x$, measured with 100 eV photons in linear polarization. (c) Schematic description of the Fermi surface nesting in $LiFe_{0.83}Co_{0.17}As$ and $NaFe_{0.955}Co_{0.045}As$. (d) Temperature dependence of $1/(^{75}T_1T)$ in LiFeAs and $LiFe_{0.83}Co_{0.17}As$. The solid lines show the fitting results to $1/(T_1T) = A + B/(T-\Theta)$.[47]

### 7.2. Local spin fluctuations in iron-based superconductors

In iron pnictides, the dc susceptibility and Knight shift are observed to increase with temperature in the normal state.[36,48–50] This behavior is always attributed to a pseudogap behavior with an unclear origin, in parallel to the underdoped cuprates.[36] From ARPES, no such pseudogap is iden-tified, which questions a band gap effect. The $A_yFe_{2-x}Se_2$ offers an ideal opportunity to figure out the nature for the thermally enhanced susceptibility. In this subsection, we review the NMR study of high-temperature spin fluctuation behavior in iron-based materials.[51]

#### 7.2.1. Absence of low-energy spin fluctuations in $A_yFe_{2-x}Se_2$ ($A$ = K, Rb, Cs,...)

The newly discovered $A_yFe_{2-x}Se_2$ system with $T_C \sim$ 30 K is an example for the deviation from the close relationship between low-energy spin fluctuations and high-$T_C$ SC in the paramagnetic phase identified by NMR.[50] As shown in Fig. 7 ($K_yFe_{2-x}Se_2$) and Fig. 30 ($Tl_{0.47}Rb_{0.34}Fe_{1.63}Se_2$), $1/(T_1T)$ in the superconducting phase shows a quadratic temperature dependence instead of the strong upturn behavior from room temperature to $T_C$, indicating the absence of low-energy spin fluctuations in the $A_yFe_{2-x}Se_2$ system.[19,22] The Fermi liquid behavior in the low-temperature limit is consistent with the case in the absence of hole pocket in the center of Brillouin zone.[21] As a result, the low-energy spin fluctuation due to interband scattering seems not to be the necessary condition for high-$T_C$ SC.

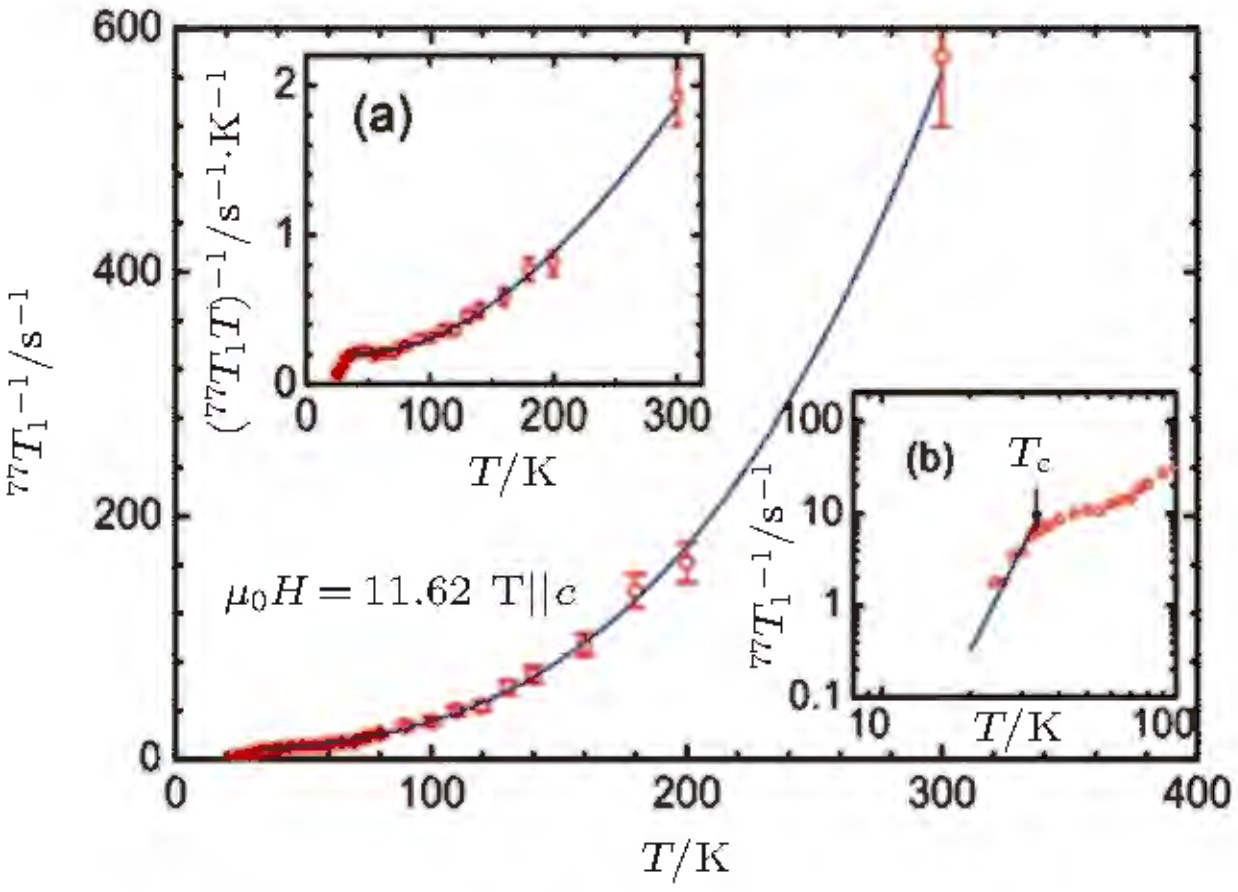

**Fig. 30.** The $1/^{77}T_1$ with an 11.62-T field applied along the $c$ axis of $Tl_{0.47}Rb_{0.34}Fe_{1.63}Se_2$ single crystals. Insets: (a) the temperature dependence of $1/(^{77}T_1T)$; (b) the $1/(^{77}T_1)$ data in the low-temperature region with a log–log scale. The solid lines are guides for the eyes.[22]

#### 7.2.2. Local spin fluctuations in $A_yFe_{2-x}Se_2$ ($A$ = K, Rb, Cs, ...)

The absence of low-energy spin fluctuations in $A_yFe_{2-x}Se_2$ has offered a valuable opportunity for studying the fluctuations from local moments. Figure 31 displays the temperature dependences of Knight shift and $1/(T_1T)$ in $Tl_{0.47}Rb_{0.34}Fe_{1.63}Se_2$ single crystals for the temperature region of 2 K $\leq T \leq$ 500 K. In order to compare the temperature dependence of Knight shift with the thermal activated behavior, Ma *et al.* plotted $\ln(K(T) - K_0)$ as a function of $1/T$ in Fig. 31(a), which clearly shows the deviation from this form below $T$ = 150 K. In fact, this supplies the evidence for the failure of the band effect for the pseudogap-like behavior. However, the Knight shift and $1/(T_1T)$ of $^{77}$Se and $^{87}$Rb each show a perfect quadratic temperature dependence, as shown in Fig. 32(b), from $T_C$ to 300 K.

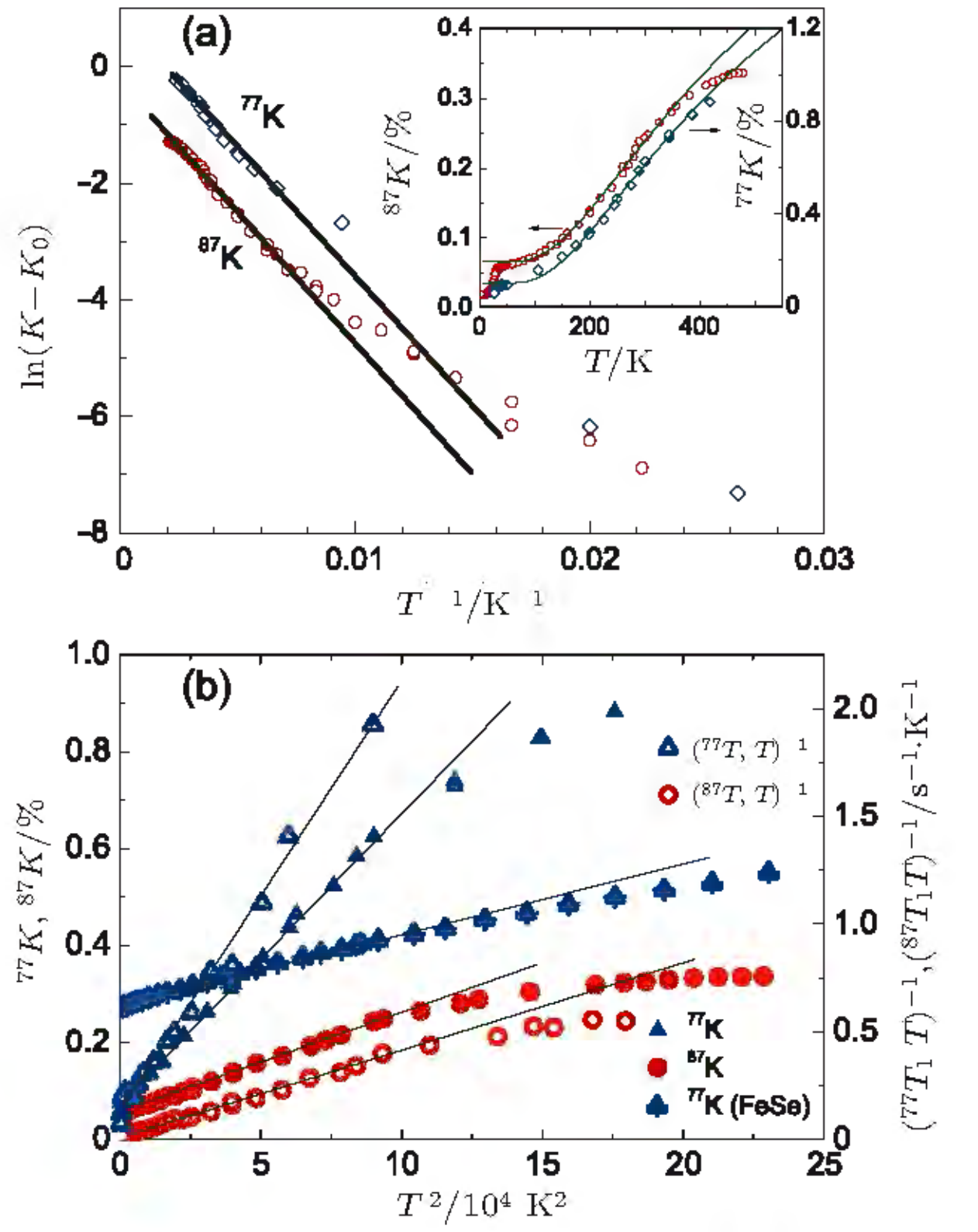

**Fig. 31.** (a) Plots of $\ln[K(T) - K_0]$ against $1/T$ for the $^{87}$Rb and $^{77}$Se nuclei in $Tl_{0.47}Rb_{0.34}Fe_{1.63}Se_2$ single crystals. The inset shows the temperature dependences of $^{87}$Rb and $^{77}$Se Knight shift. Solid lines denote the fitting results to the thermal activation function form. (b) Knight shifts and relaxation rates versus $T^2$ for both nuclei.[51] The $^{77}$K data for FeSe are cited from Ref. [37].

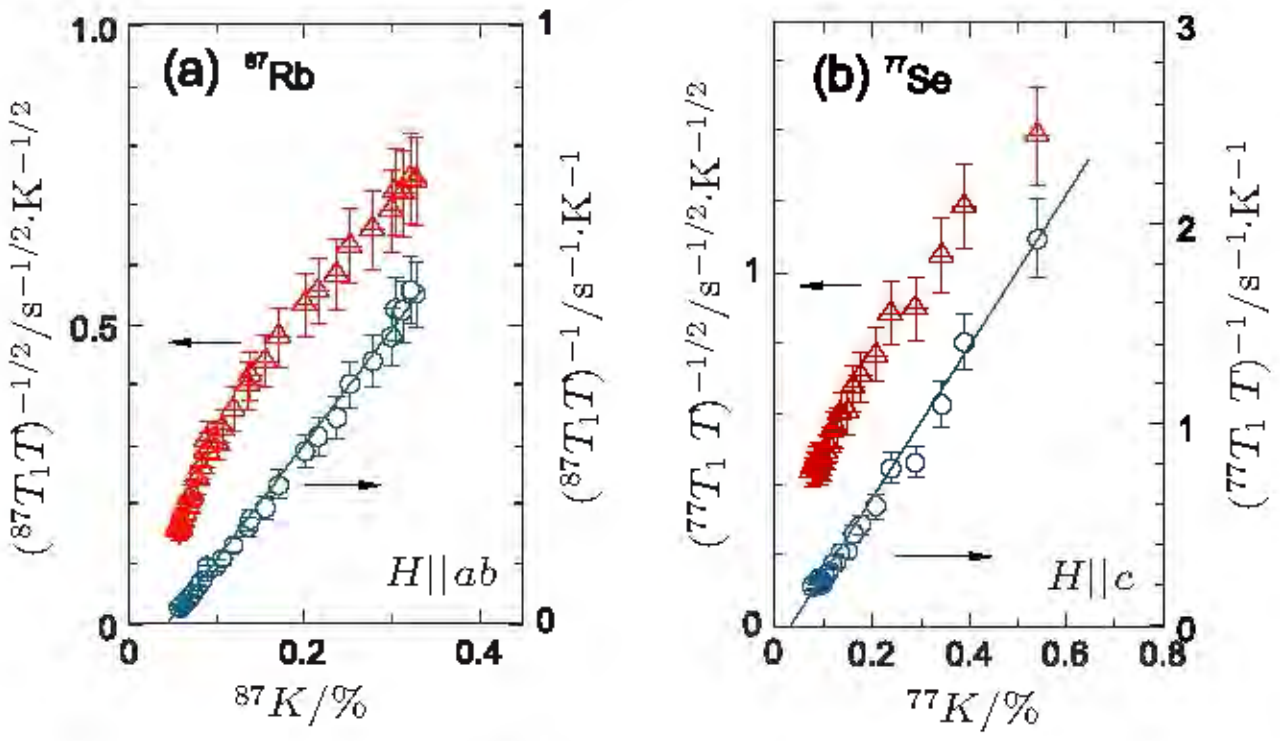

**Fig. 32.** Plots of $(T_1T)^{-1/2}$ (left-hand axis) and $(T_1T)^{-1}$ (right-hand axis) versus $K$ with the temperature as the implicate parameter for (a) $^{87}$Rb nuclei with $T_C \leq T \leq$ 425 K and (b) $^{77}$Se with $T_C \leq T \leq$ 300 K.[51]

In Fig. 32, Ma *et al.*[51] reported on $(T_1T)^{-1/2}$ and $(T_1T)^{-1}$ as a function of Knight shift to further reveal the physical meanings behind the data. As shown in Figs. 32(a) and 32(b), the system deviates from Fermi-liquid behavior with the Korringa relation, $(T_1T)^{-1/2} \propto K(T)$. Instead, $K(T)$ and $(T_1T)^{-1}$ each follow a linear relation, which suggests the thermally enhanced spin fluctuations. In particular, $K(T)$ measures spin susceptibility at $\boldsymbol{q}$=0, and $1/(T_1T)$ measures spin susceptibility at all values of $q$. Such a linear relation further suggests that the spin fluctuation exists at all values of $q$, and therefore a local spin fluctuation behavior occurs.

This linear relation with finite intercept actually suggests a two-component model for the NMR response of the paramagnetic phase. The Knight shift and $1/(T_1T)$ can be written as $K(T) = K_0 + f(T)$ and $1/(T_1T) = (1/T_1T)_0 + f'(T)$, where $K_0$ and $(1/T_1T)_0$ are the contributions from the itinerate electrons and the temperature-dependent parts, and $f(T)$ and $f'(T)$ are from the local spin fluctuations. The finite intercept actually is an indication for the existence of weakly coupled itinerate electrons in this system. The itinerate electrons contribute to the constant Knight shift and $1/(T_1T)$, and superconduct below $T_C$. The temperature-dependent part is from the spin fluctuations of local moments on Fe sites with short range order. The quadratic temperature dependence of $K(T)$ is closely related to the dimensionality of the studied system. Theoretical studies indicate that the spin fluctuations of local moments with short range order will result in a linear temperature dependence of susceptibility in two-dimensional systems, and quadratic temperature dependence in three-dimensional systems.[52] The small value of $^{77}K(T)/^{87}K(T)$, suggesting the strong magnetic interlayer coupling in $A_yFe_{2-x}Se_2$, is fully consistent with the quadratic $T$-dependence of $K(T)$.

### 7.2.3. Local spin fluctuations in $BaFe_2As_2$

These considerations about local spin fluctuations are entirely applicable for other iron-based superconductors. Figure 33 displays the high-temperature Knight shift data for $BaFe_2As_2$ parent compound. With the increase in temperature, $K(T)$ shows strict quadratic $T$ dependence, which is consistent with the three-dimensional behavior of $BaFe_2As_2$ system. For the LaFeAsO and NaFeAs systems,[53–56] the linear $T$ dependence of Knight shift again supports this local spin fluctuation scenario in iron pnictides. The local spin fluctuation suggests that the system has short-range magnetic correlation, which could also be important for high-$T_C$ SC in iron pnictides.

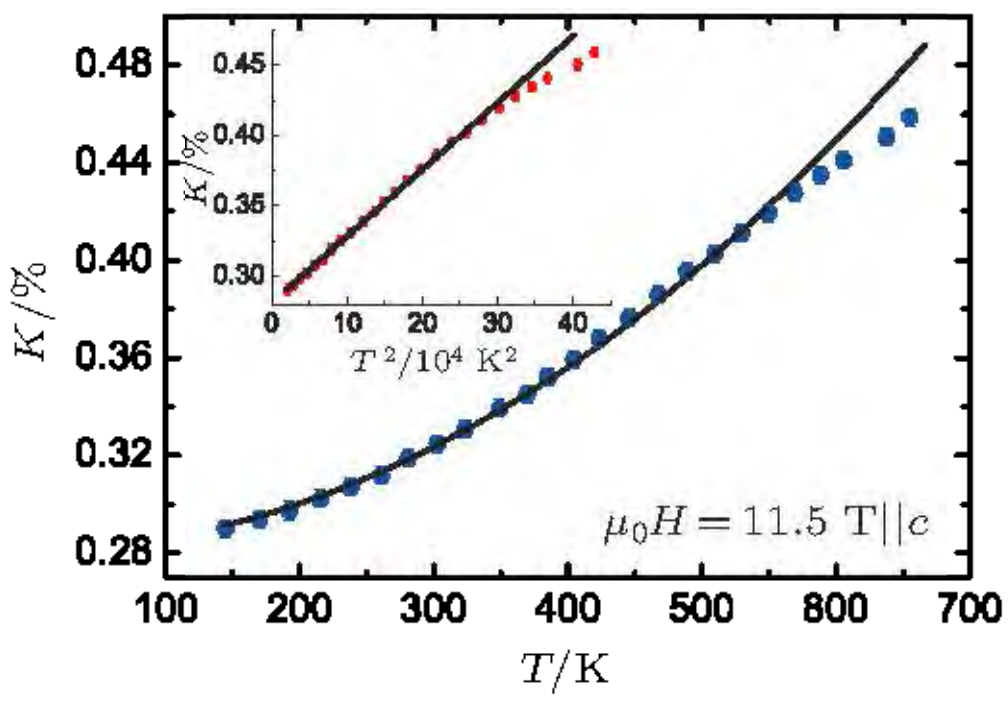

**Fig. 33.** $T$ dependence of $^{75}$As Knight shift in the $BaFe_2As_2$ parent compound for the high-temperature region. Inset: the Knight shift as a function of $T^2$.

All the experimental results suggest that there are low-energy spin fluctuations and local spin fluctuations in iron pnictides. The combination of both spin fluctuation modes may be important for the superconducting paring.

## 8. Summary

In this article, we review the NMR studies of iron-based superconductors, regarding the superconducting properties, structural and antiferromagnetic transition, the interplay between SC and AFM, as well as the spin fluctuation property. These results are summarized as follows.

(i) The paring symmetry of superconductivity. Spin singlet paring of electrons is evidenced by the sharp drop of Knight shift below $T_C$. The coherence peak is absent in the spin-lattice relaxation rate below $T_C$, which supports the unconventional SC in iron-based materials. The temperature dependence of $1/T_1$ is consistent with the $s^{\pm}$ gap symmetry with impurity scattering. However, strong nodal excitations are observed below $T_C$ in the heavily electron-doped $KFe_2As_2$ and $BaFe_2(As_{1-x}P_x)_2$, which is consistent with the low-temperature heat transport measurements. Later ARPES measurement confirms the accidental nodes in $BaFe_2(As_{1-x}P_x)_2$ with the same $s^{\pm}$ paring symmetry.

(ii) The correlations between the structural transition and antiferromagnetism. The structural transition is strongly coupled to the magnetism as evidenced by the enhanced low-energy spin fluctuations in the low-temperature orthorhombic phase.

(iii) The long-range ordered antiferromagnetism. The unconventional behaviors of the line width and magnetic moments in NaFeAs below $T_N$ are interpreted by the thermally activated domain walls. The magnetic transition temperature and magnetic moments are strongly increased in NaFeAs under high pressure, which actually stresses the importance of interlayer coupling $J_c$ in the magnetism.

(iv) The coexistence and competition between long-range antiferromagnetism and superconductivity. Microscopic coexistence between LROAFM and SC is widely observed in the underdoped $BaFe_2As_2$ system crystallized with the 122 structure.

(v) Normal state spin fluctuations. i) Low-energy spin fluctuations of stripe pattern are widely observed in iron pnictides with Fermi surface nesting. In the isoelectron-doped case, the quantum critical points are observed, which are located near and on the optimal doping level, respectively. With the doping level increasing beyond the doping level, this spin fluctuation as well as SC is strongly suppressed, supporting that the spin fluctuation is a candidate paring mechanism of superconductivty. However, for the hole-doped case, the low-energy spin fluctuation still exists in the heavily overdoped region, which supports a magnetic origin of the SC and also proves that the more extended SC dome in the hole-doped case is due to the existence of low-energy spin fluctuations. The low-energy spin fluctuations observed in $Li_{1+x}FeAs$ by NMR demonstrate that LiFeAs is also a strongly correlated system close to a magnetic ordering. The combined ARPES and NMR study on the $AFe_{1-x}Co_xAs$ crystallized into 111 structure suggests an orbital selective relation between the nesting of itinerate electrons and SC, where the nesting between the $d_{xz}/d_{yz}$ pockets, rather than the $d_{xy}$ pockets, is important for SC in iron pnictides. ii) High-temperature NMR studies indicate that the thermally activated susceptibility results from the local spin fluctuations as observed in $A_yFe_{2-x}Se_2$ ($A$ = K, Rb, Cs, ...) and $BaFe_2As_2$. The combination of the low-energy spin fluctuations and the local spin fluctuations may be important for the SC in iron-based superconductors.